\documentclass[%
 reprint,
 amsmath,amssymb,
 aps,
prd,
floatfix,
]{revtex4-2}

\usepackage{placeins}
\usepackage{graphicx}
\usepackage{dcolumn}
\usepackage{bm}
\usepackage{hyperref}

\usepackage{amsmath}
\usepackage{amssymb}
\usepackage{amsfonts}
\usepackage{accents}
\usepackage{algorithm}
\usepackage{algpseudocode}
\usepackage{float}



\def\g{\gamma}

\def\D{\Delta}
\def\p{\partial}

\def\e4c{e^{-4\chi}}
\def\ep4c{e^{4\chi}}

\def\lp{\left}
\def\rp{\right}

\begin{document}

\preprint{APS/123-QED}

\title{Stability of Non-Minimally Coupled Topological-Defect Boson Stars}

\author{Gray D. Reid}
\affiliation{Department of Physics and Astronomy,
     University of British Columbia,
     Vancouver BC, V6T 1Z1 Canada}

\author{Matthew W. Choptuik}
\affiliation{
     Department of Physics and Astronomy,
     University of British Columbia,
     Vancouver BC, V6T 1Z1 Canada}

\date{\today}

\begin{abstract}
As shown by Marunovic and 
Murkovic~\cite{black_hole_mimiker},
d-stars, composite structures consisting of a boson star and a global monopole 
non-minimally coupled to the general relativistic field, can have extremely 
high gravitational compactness. In a previous paper we demonstrated that these 
ground-state stationary solutions are sometimes additionally characterized 
by shells of bosonic matter located far from the center of 
symmetry~\cite{gdreid_2016}
stability posed by Marunovic and Murkovic, we investigate the stability of 
several families of d-stars using both numerical simulations and linear 
perturbation theory. For all families investigated, we find that the most 
highly compact solutions, along with those solutions exhibiting shells of 
bosonic matter, are unstable to radial perturbations and are therefore poor 
candidates for astrophysically-relevant black hole mimickers or other highly 
compact stable objects.

\end{abstract}

\maketitle

\section{Background\label{bsgm_l_sec_background}}
Attempts to create stable solitonic solutions in the context of general 
relativity go back to  Wheeler in 1955 with the development of 
geons--solitonic objects comprised of various fundamental fields coupled to 
gravity~\cite{wheeler_geons}. Although Wheeler's geons proved to be unstable 
in general, further work by Kaup~\cite{kaup_klein_gordon_geon} and Ruffini 
and Bonazzola~\cite{ruffini_1969_systems_of_self_gravitating_particles},
lead to the discovery of the stable massive solitons today known as boson 
stars \cite{liebling2012dynamical}.

Over the intervening years, boson stars and their descendants  have been 
invoked for a large variety of processes and models including black hole 
mimickers~\cite{barranco_2011_constraining_scalar_black_hole, 
black_hole_mimiker}, models of neutron stars~\cite{torres_2000_supermassive, 
yuan2004constraining, palenzuela_2008_orbital_dynamics, 
palenzuela_2007_collisions_of_boson_stars, black_hole_mimiker}, binary 
systems~\cite{palenzuela_2008_orbital_dynamics}, sources of dark 
matter~\cite{mielke1999boson, schunck2000boson, nucamendi2001alternative, 
Lopez2010_bosonic_gas, rindler2012angular, khlopov1985gravitational} 
and sources of gravitational 
waves~\cite{mielke1999boson, palenzuela_2008_orbital_dynamics}. Though boson
stars are not known to exist in nature, the simplicity of their matter model 
makes them a valuable tool for qualitative analysis and for providing a simple 
first step and test bed for more complex matter 
models~\cite{liebling2012dynamical, palenzuela_2008_orbital_dynamics}.

Studies have demonstrated that boson stars are stable to perturbations 
provided that the central density of the star is sufficiently 
small~\cite{lee1989stability, gleiser1989gravitational, 
kusmartsev1991gravitational,  kusmartsev1991gravitational, mielke1999boson, 
liebling2012dynamical}. However, without a self-interaction term in the
potential, the mass of the star scales as $\frac{1}{2}m_p^2/{m}$ (where 
$m_p$ is the Planck mass and $m$ is the boson mass). 
For reasonable particle masses, this results in stars with masses far below 
the usual Chandrasekhar limit for fluid stars~\cite{colpi_boson_stars_1986, 
liebling2012dynamical, amaro2010constraining}. Thus, these so-called 
mini-boson stars are useful primarily as a test bed with their more 
specialized cousins (having, for example, additional terms in the potential) 
being adapted to various astrophysical 
situations~\cite{liebling2012dynamical}.

Whereas boson stars gain their stability through a conserved charge and the 
interplay between pressure and gravity, global monopoles are topologically 
stable~\cite{barriola1989gravitational, shi1991gravitational, 
vilenkin2000cosmic}. Along with other topological defects such as textures, 
domain walls, and strings, monopoles are expected to form fairly generically 
when underlying field symmetries are broken through early-universe phase 
transitions which are mediated by expansion and 
cooling~\cite{vilenkin2000cosmic}.

In the case of the monopole, the simplest class of defect consists of 
a scalar field triplet with a global $O(3)$ symmetry which is 
spontaneously broken to $U(1)$ on a non-contractible 2-surface. If 
the broken symmetry is local, the resulting monopole is shielded by the 
Maxwell field and has finite energy and extent. Conversely, if the field 
exhibits a global symmetry, we find that the resulting energy is linearly 
divergent in radius~\cite{barriola1989gravitational, shi1991gravitational, 
vilenkin2000cosmic, ADM_mass}.

Although this divergence may seem somewhat problematic, there are two important 
caveats. First, the energy divergence cuts off upon encountering 
another monopole or antimonopole. Second, in the context of general 
relativity, the energy divergence has the simple effect of producing a solid 
angle deficit spacetime along with a small effective negative mass core,
rather than more exotic features~\cite{gdreid_2016, ADM_mass, 
shi1991gravitational, barriola1989gravitational}. As shown by 
Barriola et al.~\cite{barriola1989gravitational}, we would expect global 
monopoles and anti-monopoles to annihilate extremely efficiently due to the 
fact that the interaction strength between them is independent of distance. 
Although this annihilation is avoided by local monopoles, we expect 
a Hubble volume to contain only $\approx1$ global monopole at the present 
time due to the efficiency of this interaction. 

Putting aside these considerations, when a global monopole and boson star 
are combined, the result is a novel object referred to as a topological-defect 
boson star or ``d-star". Previously studied 
in~\cite{li1995fermion, li2000boson, black_hole_mimiker,gdreid_2016}, 
it was shown that through non-minimal coupling and proper choice of 
interaction parameters, d-stars could be made extremely dense, thereby 
potentially acting as mimickers of black holes or other highly compact 
objects~\cite{black_hole_mimiker}. 
Subsequent in-depth investigation of these objects revealed novel interactions 
and ground state solutions~\cite{gdreid_2016}. When viewed as functions of the 
boson star central density, these ground state solutions are characterized by 
discontinuous changes in the global properties of the system (mass, charge,
etc.).
To better describe this behaviour, we borrow the terminology of  
statistical mechanics.  In this analogy, the central density of the boson 
star takes the place of the temperature, the asymptotic mass takes the role 
of the energy and the mass gap is similar to latent heat.

The discontinuous changes in global properties are mediated by the 
appearance or disappearance of shells of bosonic matter at characteristic 
radii which can be either finite or infinite. We use the term \em asymptotic 
shell \em to refer to any shell of matter which first appears far from the 
coordinate origin as $\psi(0)$ is increased past some critical value and 
which subsequently vanishes when $\psi(0)$ is further increased past a second
critical value. 
We refer to those families of solutions with  mass gaps (when the mass is 
viewed as a function of the central density) as expressing a first order 
phase transition. Those with discontinuities in the derivative of the 
asymptotic mass or charge express a second order phase transition. The 
interested reader is directed to~\cite{gdreid_2016} for an in-depth review. 

\section{Review of Stationary D-Stars\label{bsgm_l_sec_review}}

We have previously solved the equations of motion assuming stationary 
solutions, the harmonic ansatz for the boson field and a hedgehog ansatz for 
the monopole fields~\cite{gdreid_2016}. The solutions we discovered were 
characterized by a series of discrete boson star central amplitudes, 
$\psi_i^c(0)$, about which the character of the solutions changed 
discontinuously in a manner analogous to a phase transition. In what 
follows, we will use the same terminology and notation as our previous 
paper~\cite{gdreid_2016}, which is briefly reviewed below. 

\begin{figure}[!htb]
\centering
\includegraphics[scale=1.0]
{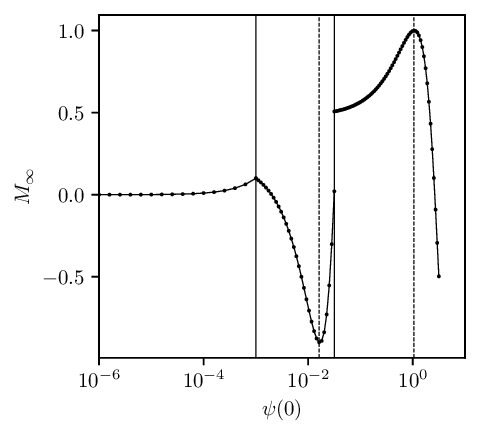}
\caption{Asymptotic mass as a function of central amplitude for a hypothetical 
family consisting of three branches. The first branch consists of a single 
region while each of the subsequent branches consist of two regions, the 
extent of which are delimited by their mass turning points. Solid vertical 
lines denote the extent of branches while regions within a branch are 
separated with vertical dashed lines. }
\label{bsgm_l_mass_plot_example}
\end{figure}

The parameter space we consider here is six-dimensional, spanned by the 
central amplitude of the boson star, $\psi(0)$, and five coupling parameters: 
the solid angle defect, $\Delta^2$, the quartic global monopole potential 
parameter, $\lambda_G$, the quartic boson star potential parameter, 
$\lambda_B$, the global monopole non-minimal coupling, $\xi_G$, and the boson 
star non-minimal coupling, $\xi_B$. We fix the mass of the boson star field, 
$m=1$, and note that this sets the energy scale of the solutions. We define 
a \em family \em of solutions to be the set of all ground state solutions with 
common $\Delta$, $\lambda_G$, $\lambda_B$, $\xi_G$ and $\xi_B$. As such, 
within a given family, solutions can be indexed by the boson star central 
amplitude, $\psi(0)$, which is the only free parameter of the family (see 
Fig.~\ref{bsgm_l_mass_plot_example}).

Due to the large parameter space associated with these solutions, it was not 
feasible to perform a comprehensive parameter space survey. Instead, as 
in~\cite{gdreid_2016}, we focus on a number of families of solutions which 
appear to capture the novel behavior associated with the model. In 
subsequent sections we deal with eight families of solutions whose fixed 
parameters are given in Table~\ref{bsgm_l_table_families}. For simplicity, 
the boson star quartic self interaction coupling constant, $\lambda_{B}$, has 
been set to 0 under the assumption that its primary effect will be to produce 
more compact objects (while having only a minor effect on the low density 
asymptotic shells that we find). Those families that were investigated 
in~\cite{gdreid_2016} were given identical designations.

\begin{table}[!ht]
\centering
\begin{ruledtabular}
\begin{tabular}{ c  c  c  c  c  c }
    Family & $\Delta^2$ & $\lambda_{B}$ & $\lambda_{G}$ & $\xi_{B}$ 
    & $\xi_{G}$
    \\ 
    \hline
    $c$   & 0.36 & 0    & 1.000        & 0        & 0        \\  
    $d$   & 0.81 & 0    & 0.010        & 0        & 0        \\  
    $e$   & 0.25 & 0    & 0.001        & 3        & 3        \\  
    $f$   & 0.49 & 0    & 0.010        & 5        & 0        \\
    $g$   & 0.09 & 0    & 0.010        & 0        & 5        \\
    $h$   & 0.08 & 0    & 0.100        & $-4$     & 5        \\
    $p_1$ & 0.09 & 0    & 0.040        & 0        & 0        \\
    $p_2$ & 0.25 & 0    & 0.040        & 0        & 0        \\  
\end{tabular}
\end{ruledtabular}
\caption{Families of solutions and their associated parameters. Each family 
consists of a continuum of solutions labelled by the central amplitude 
of the boson star. In particular, family $h$ corresponds to a family in the 
high compactness regime as defined in~\cite{black_hole_mimiker}. Due to their 
relatively simple and illustrative modal structure, families $p_1$ and $p_2$ 
are the only ones we explore with perturbation theory. 
}
\label{bsgm_l_table_families}
\end{table}

We define a \textit{branch} of a family to be the set of all solutions
in the family where the asymptotic mass, $M_\infty$, is $C^1$ as a function 
of the central amplitude, $\psi(0)$. Using this definition, 
Fig.~\ref{bsgm_l_mass_plot_example} provides a mass plot illustrating a 
hypothetical family with three branches. We use the term \textit{region} to 
refer to the set of all solutions on a given branch between extremal values 
of the asymptotic mass. Using our previous example, the first branch of 
Fig.~\ref{bsgm_l_mass_plot_example} consists of a single region while the 
second and third branches each consist of two regions.

\begin{figure}[!htb]
\centering
\includegraphics[scale=1.0]
{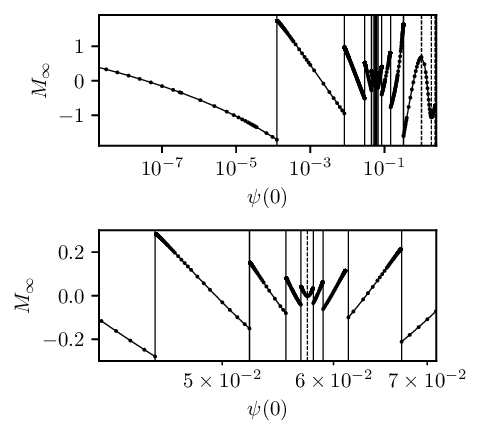}
\caption{Asymptotic mass as a function of central amplitude for family $d$. The 
lower panel shows an expanded view of the upper plot highlighting the central 
structure. As demonstrated in~\cite{gdreid_2016}, the apparent discontinuities 
are genuine. These discontinuities correspond to shells of bosonic matter of 
finite mass and particle number appearing or disappearing at spatial infinity. 
}
\label{bsgm_l_mass_plot_5}
\end{figure}

As demonstrated in Fig.~\ref{bsgm_l_mass_plot_5}, which plots the asymptotic 
mass, $M_{\infty}$, of a family of solutions, this mass parameter is not in 
general a smooth function of the boson star central amplitude, $\psi(0)$, as 
would be expected for a fluid star. As we construct families of stationary 
solutions by varying the central amplitude of the boson star (keeping 
all other parameters fixed) we find that when the central amplitude is 
increased or decreased across a critical point $\psi^c_i(0)$, a shell of 
bosonic matter will either appear or vanish far from the center of symmetry. 
As shown in~\cite{gdreid_2016}, these shells of matter may either appear 
suddenly at spatial infinity, or gradually at a finite radius when the boson 
star is non-minimally coupled to gravity.

As discussed in full detail in~\cite{gdreid_2016}, these families of 
solutions have features that are in many ways analogous to critical points and 
phase transitions in thermodynamical systems. When shells appear at infinity 
with finite mass, we have a direct analogy with first order phase transitions 
with $\psi(0)$ taking the role of the temperature or pressure and the mass gap 
being analogous to the latent heat. When the transition is gradual, as in the 
case of non-minimal coupling, we have a situation more analogous to second 
order, or continuous, phase transitions. 

\section{Overview\label{bsgm_l_sec_overview}}

In analysing the stability of the boson d-star solutions, we adopt a two 
pronged approach. First, we consider the general non-minimally coupled case 
and perform dynamical simulations of a number of families which seem to be 
representative of the model as a whole. Specifically, from these families we 
choose a few solutions from each branch, perturb the solutions and follow 
the evolution of the system, looking for growth of excited modes. 
Previous studies have shown that stability transitions are confined to turning 
points of the asymptotic mass or charge when these quantities are viewed as 
functions of boson star central amplitude~\cite{lee1989stability, 
kleihaus2012stable, gleiser1989gravitational, kusmartsev1991gravitational}. We 
greatly simplify our investigation by considering only a small number of 
evolutions per region and by assuming that the observed stability 
for these simulations generalizes across the entire 
region.  

Our second approach involves a detailed analysis of the mode structure of 
the d-star solutions via linear perturbation analysis. Due to the complexity 
of the resulting equations for the non-minimally coupled case, we limit 
ourselves to the investigation of the minimally coupled configurations. 
Through an exhaustive investigation of two families, we deduce a general 
mode structure which is in broad agreement with the results of our dynamical 
simulations.

\section{Matter Model\label{bsgm_l_sec_matter_model}}

Following the prescription of Marunovic and Murkovic 
\cite{black_hole_mimiker}, the dimensionless Einstein-Hilbert action ($c=1, 
G=1/8\pi$) is given by
\begin{equation}
   S_{\mathrm{EH}}=\int{dx^4\sqrt{-g}\,\frac R2}
\end{equation}
while the actions for the boson star and global monopole are: 
\begin{align}
   \begin{split}
   S_{B}&=\int{}dx^4\sqrt{-g}\left[-\frac{1}{2}\left(\nabla_u{\Psi^{*}}
   \right)\left(\nabla^u
   {\Psi}\right)-V_{B} 
   \right.
   \\
   &\mathopen{}\left.  \hphantom{=}
   +\frac{\xi_{B}}{2}R\left(\Psi^*\Psi\right)\right],
\end{split}
\\
\begin{split}
   S_{G}&=\int{}dx^4\sqrt{-g}\left[-\frac{\Delta^2}{2}\left(\nabla_u{\phi^a}
   \right)\left(\nabla^u
   {\phi^a}\right)-V_{G}
   \right.
   \\
   &\mathopen{}\left.  \hphantom{=}
   +\frac{\xi_{G}}{2}R\Delta^2\left(\phi^a\phi^a\right)\right] .
\end{split}
\end{align}
Here, $\Psi$ is the complex boson star field,  $\phi^a$ are scalar field 
triplets comprising the monopole, $V_{B}$ and $V_{G}$ are the self interaction 
potentials for the boson field and monopole fields respectively, $R$ is the 
Ricci scalar and $\xi_{B}$ and $\xi_{G}$ are the non-minimal coupling 
constants. The stress-energy tensors associated with the matter actions are:
\begin{align}
\label{bsgm_l_T_uv_B}
\begin{split}
   T^{B}_{\mu\nu}&=\frac{1}{2}\nabla_{\mu}\Psi^{*}\nabla_{\nu}\Psi
   +\frac{1}{2}\nabla_{\nu}\Psi^{*}\nabla_{\mu}\Psi        
   \\
   & \hphantom{=}
   -\frac{1}{2}g_{\mu\nu}\left(\nabla_{\alpha}\Psi 
   \nabla^{\alpha}\Psi^{*} +2V_{B} \right)  
   \\
   & \hphantom{=}
   -\xi_{B}\left(G_{\mu\nu} +g_{\mu\nu}\nabla_\alpha \nabla^{\alpha}
   - \nabla_{\mu} \nabla_{\nu}\right)\Psi\Psi^{*},
\end{split}
\end{align}
\begin{align}
\label{bsgm_l_T_uv_G}
\begin{split}
   T^{G}_{\mu\nu}&=\frac{\Delta^2}{2}\nabla_{\mu}\phi^{a}\nabla_{\nu}\phi^{a}
   +\frac{\Delta^2}{2}\nabla_{\nu}\phi^{a}\nabla_{\mu}\phi^{a}
   \\
   & \hphantom{=}
   -\frac{1}{2}g_{\mu\nu}\left(\Delta^2\nabla_{\alpha}\phi^{a}
   \nabla^{\alpha}\phi^{a} +2V_{G} \right) 
   \\
   & \hphantom{=}
   -\xi_{G}\Delta^2\left(G_{\mu\nu} +g_{\mu\nu}\nabla_\alpha
   \nabla^{\alpha}- \nabla_{\mu} \nabla_{\nu}\right) \phi^a\phi^a.
\end{split}
\end{align}

We use the standard 3+1 decomposition where 4D spacetime is foliated into a 
sequence of spacelike hypersurfaces, $\Sigma_t$, such that each hypersurface 
of constant $t$ has a 3-metric, $\gamma_{ij}$. Explicitly, the 4-metric takes 
the form
\begin{align}
   g_{\mu \nu}=\left(
   \begin{array}{cc}
   -\alpha^2+\beta^i\beta_i & \beta_j \\
   \beta_i & \gamma_{ij} \\
   \end{array}
   \right).
\end{align}
with a time-like normal, $n^\nu$, to the foliation, $\Sigma_t$, given by
\begin{align}
   n^\nu = \lp(\frac{1}{\alpha},-\frac{\beta^i}{\alpha}\rp).
\end{align}
Here, $\alpha$ and $\beta^i$ are the usual lapse and shift respectively. We 
impose spherical symmetry, and adopt polar-areal coordinates such that the 
line element becomes
\begin{align}
   ds^2&=-\alpha(t,r)^2dt^2+a(t,r)^2dr^2
   \\ \nonumber
   & \hphantom{=}
   +r^2 \lp(d\theta^2 
   + \sin^2\theta d\phi^2\rp).
\end{align}

Decomposing the boson field into a real and imaginary part and taking the 
hedgehog ansatz for the monopole, the matter fields and corresponding 
potentials are
\begin{align}
   \Psi&=\phi_R + i\phi_I,
   \\
   \label{bsgm_l_hedgehog}
   \phi^a&=\phi_M\frac{x^a}{r},
   \\
   \label{bsgm_l_V_G}
   V_{G}&=\frac{\lambda_{G}}{4}\Delta^4(\phi_M^2-1)^2 ,
   \\
   V_{B}&=\frac{m^2}{2}\left(\phi_R^2+\phi_I^2\right) +
   \frac{\lambda_{B}}{4}\left(\phi_R^2+\phi_I^2\right)^2.
\end{align}
Varying the actions with respect to the matter fields gives the equations of 
motion for the matter:
\begin{align}
   \nabla_\mu \nabla^{\mu} \phi_R&=\phi_R\partial V_{\phi_R} - \xi_{B}R\phi_R,
   \\
   \nabla_\mu \nabla^{\mu} \phi_I&=\phi_I\partial V_{\phi_I} - \xi_{B}R\phi_I,
   \\
   \nabla_\mu \nabla^{\mu} \phi_M&=\frac{\phi\partial V_{\phi_M}}{\Delta}
   +\frac{2\phi_M}{r^2}- \xi_{G}R\phi_M,
\end{align}
where:
\begin{align}
   \partial V_{\phi_R} &= m^2\phi_R + \lambda_{B}\left(\phi_R^2+\phi_I^2
   \right)\phi_R,
   \\
   \partial V_{\phi_I} &= m^2\phi_I + \lambda_{B}\left(\phi_R^2+\phi_I^2
   \right)\phi_I,
   \\
   \partial V_{\phi_M} &= \lambda_{G}\Delta^4\left(\phi_M^2-1\right)\phi_M.
\end{align}
We express the fields $\phi_A = (\phi_R, \phi_I, \phi_M)$ in terms of their 
conjugate momentum and spatial derivatives:
\begin{align}
   \label{bsgm_l_matter_start}
   \Pi_A &= \frac{a}{\alpha}\p_t \phi_A
\\
   \Phi_A &= \p_r \phi_A
\end{align}
Evaluating tensor components and simplifying, the equations of motion for the 
matter fields may be expressed as
\begin{align}
   \begin{split}
   \partial_t \Pi_R&=-{\left(\xi_{B}\phi_R T+\partial_{\phi_R} V\right)\alpha
   a}+\p_r\left( \frac{\Phi_R \alpha}{a} \right)
   \\
   & \hphantom{=}
   + \frac{2\Phi_R \alpha}{r a},
   \end{split}
\\
   \begin{split}
   \partial_t \Pi_I&=-{\left(\xi_{B}\phi_I T+\partial_{\phi_I} V\right)\alpha
   a}+\p_r\left( \frac{\Phi_I \alpha}{a} \right) 
   \\
   & \hphantom{=}
   + \frac{2\Phi_I \alpha}{r a},
   \end{split}
\\
   \begin{split} 
   \partial_t \Pi_M&=-{\left(\xi_{G}\phi_MT+
   \frac{\partial_{\phi_M} V}{\Delta^2}+\frac{2\phi_M}
   {r^2}\right)\alpha a}
   \\
   & \hphantom{=}
   +\p_r\left( \frac{\Phi_M \alpha}{a} \right) +\frac{2\Phi_M\alpha}{ar}.
   \end{split}
\\
   \begin{split}
   \partial_t \Pi_P&=\p_r\left( \frac{\Phi_P \alpha}{a} \right) 
   + \frac{2\Phi_P \alpha}{r a}.
   \end{split}
   \label{bsgm_l_matter_end}
\end{align}
Here, we have used the contracted Einstein equation $R=-T$, and have 
incorporated  a massless scalar field, $\phi_P$, to facilitate perturbation
of the stationary solutions. The choice of polar-areal coordinates greatly 
simplifies the Einstein equations and after a considerable amount of 
manipulation we arrive at the form of the equations given in 
Appendix~\ref{bsgm_l_app_eom}.

\subsection{Boundary Conditions\label{bsgm_l_sec_boundary}} 

Given the hedgehog ansatz (\ref{bsgm_l_hedgehog}), $\phi_M$ is the magnitude 
of the global monopole fields, $\phi^a$, which are analogous to an outward 
pointing vector field. As such, to maintain regularity we must have  
$\phi_M=0$, at the center of symmetry. Further, as $r\rightarrow0$, regularity 
requires $\phi_R$, $\phi_I$, $\phi_P$, $\Pi_R$, $\Pi_I$, $\Pi_P$, $a$ and 
$\alpha$ be even functions of $r$ (with $a(t,0) = 1$)  while $\phi_M$ and 
$\Pi_M$ are odd functions of $r$.

In the limit that $r \rightarrow \infty$, the boson star field  exponentially 
approaches zero while the global monopole transitions to its vacuum state 
$(\phi_R \rightarrow 0$, $\phi_I \rightarrow 0$, 
$\phi_M \rightarrow 1 + \sum_i c_i r^{-i})$. 
Defining $\tilde \Delta$ by
\begin{equation}
   \label{bsgm_l_delta_tilde}
   \tilde{\Delta}=\frac{\Delta^2}{1+\xi_{G}\Delta^2},
\end{equation}
and assuming a series expansion in $1/r$, the metric equations can be 
integrated to yield the following regularity conditions as $r$ approaches 
infinity~\cite{nucamendi2001alternative, black_hole_mimiker}:
\begin{align}
   \label{bsgm_l_bc_start}
   \phi_R &= \phi_I = \phi_P = 0,
   \\
   \phi_M &= 1-\frac{1}{\lambda_{G}
   \Delta^2r^2\left(1+\xi_{G} \Delta^2\right)},
   \\
   \Pi_R &= \Pi_I = \Pi_P = \Pi_M =0,
   \\
   \label{bsgm_l_assym_a}
   a &= \left(1-\tilde{\Delta}-
   \frac{2M}{r}\right)^{-1/2},
   \\
   \label{bsgm_l_assym_alpha}
   \alpha &= \left(1-\tilde{\Delta}-
   \frac{2M}{r}\right)^{1/2}.
\end{align}
Here, $M$ is a constant of integration proportional to the  ADM mass of a 
solid angle deficit spacetime as defined in the next section.

\subsection{Conserved and Diagnostic Quantities
\label{bsgm_l_sec_conserved_quantities}}

The global $U(1)$ invariance of the boson star field gives rise to a 
conserved current,
\begin{align}
   J_\mu&=\frac{i}{16\pi}\left(\Psi^*\nabla_\mu\Psi-\Psi\nabla_\mu\Psi^*
   \right),
   \\ \nonumber
   &=\frac{1}{8\pi}\left(\phi_I\nabla_\mu\phi_R-\phi_R\nabla_\mu\phi_I\right),
\end{align}
with temporal component,
\begin{align}
   J_t&=\frac{\alpha}{8\pi a}\left(\phi_I \Pi_R-\phi_R \Pi_I\right).
\end{align}
Associated with the current is a conserved charge,
\begin{align}
   N&=\int{J_\nu n^\nu\sqrt{\gamma}}dx^3,
   \label{bsgm_l_noether}
\end{align}
with spatial gradient,\begin{align}
   \partial_rN&={\frac{r^2}{2}\left(\phi_I \Pi_R - \phi_R \Pi_I\right)}.
\end{align}

Although the energy of the spacetime is linearly divergent in $r$, it is 
possible to use the prescription of Nucamendi et al.~\cite{ADM_mass}  to 
define an ADM-like mass, $M_{\mathrm{ADM}}$, for a solid angle deficit 
spacetime as
\begin{align}
\begin{split}
   \label{bsgm_l_M_ADM}
   M_{\mathrm{ADM}} &= \frac{1}{16\pi\lp( 1-\tilde{\Delta} \rp)}
   \int_{\partial \Sigma_t} \lp( \bar{\g}^{ac} \bar{\g}^{bd} - \bar{\g}^{ab}
   \bar{\g}^{cd} \rp)
   \\
   & \hphantom{=}
   \bar{D}_b\lp(\g_{cd}\rp) dS_a.
\end{split}
\end{align}
Here $\bar{\g}_{ab}$ is a metric which is flat everywhere save a deficit 
solid angle, and which is induced on all constant time hypersurfaces, 
$\Sigma_t$.  $\bar{D}_b$ is the associated connection and $dS_a$ is the 
surface area element~\cite{ADM_mass}.

Evaluation of~(\ref{bsgm_l_M_ADM}) using the asymptotic forms of the metric 
functions~(\ref{bsgm_l_assym_a})--(\ref{bsgm_l_assym_alpha})  under the 
coordinate changes prescribed by Nucamendi, yields
\begin{align}
M_{\mathrm{ADM}}=M{\left(1-\tilde{\D}\right)^{-3/2}},
\end{align}
where $M$ is defined as in Eqn.~(\ref{bsgm_l_assym_a}). We further 
define a mass function as
\begin{align}
        M(t,r) &= \frac{r}{2}\left(1-a(t,r)^{-2}-\tilde{\D}\right),
        \\
        M_{\infty} &\equiv \lim_{r\rightarrow\infty} M(t,r),
\end{align}
and use it to monitor energy conservation of the system. We also use the 
above definition of $M(t,r)$ to define a measure of the compactness of the 
system. Specifically, following Marunovic and 
Murkovic~\cite{black_hole_mimiker}, we define the compactness as
\begin{align}
C(t,r)\equiv \frac{2M(t,r)}{r\lp(1-\tilde{\D}\rp)},
\end{align}
such that $C(t,r_0)\rightarrow1$ indicates the development of an apparent horizon at areal 
radius $r_0$. Correspondingly, we define the quantity $C_{\mathrm{max}}$ as
\begin{align}
C_{\mathrm{max}}\equiv\max\left({C(t,r)}\right).
\end{align}
For any given configuration of matter, then, $C_{\mathrm{max}}$ measures the 
maximum compactness of the configuration.

\section{Dynamical Simulation\label{bsgm_l_sec_ds}}

This section provides an overview of our evolution scheme and 
associated numerics. Section~\ref{bsgm_l_subsec_ds_initial_data} details the 
initialization of the metric and matter fields, while 
Sec.~\ref{bsgm_l_subsec_ds_evolution} introduces the finite difference 
discretizations used to solve the equations of motion and  describes the 
evolution procedure. Sec.~\ref{bsgm_l_subsec_ds_convergence} specifies the 
tests used to ensure convergence of our scheme. Finally, 
Sec.~\ref{bsgm_l_subsec_ds_growth_modes} describes our method of extracting 
stable and unstable perturbative modes from the dynamical simulations.

\subsection{Initial Data\label{bsgm_l_subsec_ds_initial_data}}

To initialize an evolution, the boson star fields, $\phi_R$  and $\phi_I$, 
global monopole field, $\phi_M$, and metric fields, $a$ and $\alpha$, are 
interpolated to the evolution grid from a stationary solution (computed 
using the methodology described in~\cite{gdreid_2016}) using a high 
order interpolation scheme. Subsequently, we add a small perturbation 
consisting of either a Gaussian pulse in the massless scalar field, $\phi_P$, 
or a rescaling of the matter fields. Finally, the Hamiltonian and momentum 
constraints are re-integrated to account for the perturbation and the system 
is ready for evolution.

Specifically, when considering a case where the excited modes are observed to 
grow slowly compared to the light-crossing time of the star, the perturbation 
is set using a time-symmetric massless scalar field pulse of the form
\begin{align}
   \phi_P(0,r)=a_0\exp\left(-\frac{\left(r-r_0\right)^2}{\sigma_0^2}\right),
\end{align}
where $a_0$, $\sigma_0$ and $r_0$ determine the location and intensity of 
the pulse at $t=0$. 
A portion of this time symmetric pulse implodes inwards and excites 
perturbative modes in the boson d-star before dispersing to infinity. 

When the growth rate of the perturbative modes are large compared to the 
scale of the star, this approach fails to produce good results (e.g.~the 
perturbations evolve into the non-linear regime before the perturbing pulse 
is able to disperse). In this case, the  truncation error induced by 
restricting the stationary solutions to the evolution grid (which is quite 
coarse compared to the one used to determine the time-independent 
solutions) induces growth modes over which we have very little control. 
These modes quickly become the dominant source of perturbation and hamper 
the extraction of useful information from the simulations. To overcome this, 
we introduce a perturbation by rescaling the matter fields as,
\begin{align}
\phi_i(x)\rightarrow \phi_i\left(\left(1+\lambda\right) x\right),
\end{align}
where $\lambda$ is taken to be a small number, typically on the order of 
$10^{-5}$, and reintegrating the Hamiltonian and polar slicing condition.  
Here, $x$, is a compactified spatial coordinate as will be discussed shortly. 
Although this form of perturbation works well, it is decidedly less natural 
than perturbing with an external matter field and we stick to the former 
approach whenever possible. 

\subsection{Evolution Scheme\label{bsgm_l_subsec_ds_evolution}}

We evolve the matter field quantities using 
(\ref{bsgm_l_matter_start})--(\ref{bsgm_l_matter_end}) and a second order 
finite difference scheme with Crank-Nicholson differencing. To damp high 
frequency solution components we add fourth order, temporally centered,  
Kreiss-Oliger dissipation. Due to the global nature of the monopole field, 
the need to evolve the simulations for many dynamical time scales, and the 
fact that it helps in implementing the $r\rightarrow\infty$ boundary 
conditions, we adopt compactified coordinates defined by:
\begin{align}
        r &= \frac{\lambda x}{1-x}, \quad
        0 \le x \le 1 ,
\end{align}
where $\lambda$ is a positive real number  that typically satisfies 
$1\le\lambda\le100$. Specifically, $\lambda$ is chosen so that all solution 
features are well resolved with our choice of mesh spacing. Since $x$ 
compactifies the entire domain of $r$, this coordinate change works in 
conjunction with our numeric dissipation operators to suppress wave like 
oscillations far from $r=0$. This in turn permits us to forgo crafting 
outgoing boundary conditions for our fields and instead impose trivial 
boundary conditions corresponding to 
(\ref{bsgm_l_bc_start})--(\ref{bsgm_l_assym_alpha})
at the $x=1$ limit of our domain. Care must be taken, however, to ensure 
that the parity of functions at the origin is treated according to their 
behaviour in $r$ rather than $x$.

As noted in Appendix~\ref{bsgm_l_app_eom}, it is possible to find an 
expression for $\partial_r a$ independent of $\alpha$. As such, we may 
consider the equation for the metric as two initial value problems rather 
than a coupled boundary value problem. In practice, we find the most 
effective method of solving for the metric functions is to integrate 
$\partial_r a$ from $x=0$ to $x=1$, initialize 
$\alpha=1/a$ at $x=1$, and integrate $\partial_r\alpha$ back to $x=0$. 

Although our overall evolution scheme is well suited to the evolution of 
highly dynamical simulations, its utility for investigating nearly stationary 
solutions and the growth of perturbations is limited by the use of second 
order finite difference operators. In particular, there are significant 
restrictions on the period of time for which a simulation may be run before 
dispersion becomes the dominant factor limiting solution accuracy. The scheme 
was chosen for ease of implementation, but fourth order finite difference or 
spectral schemes would be far superior for the purpose of resolving modes 
with very slow growth rates.

\subsection{Convergence\label{bsgm_l_subsec_ds_convergence}}

To validate the stability of the evolution scheme and to ensure that mass and 
charge are approximately conserved for dynamic configurations, we choose 
initial data consisting of a stationary solution perturbed by a 
large-amplitude massless scalar field pulse at the origin. The mass energy 
of the massless scalar field is a significant fraction of the total mass 
energy of the system, so the system as a whole is highly perturbed from 
stationarity. Specifically, the monopole is non-stationary far from the 
origin due to its coupling to the modified metric functions.

\begin{figure}[!htb]
\centering
\includegraphics[scale=1.0]
{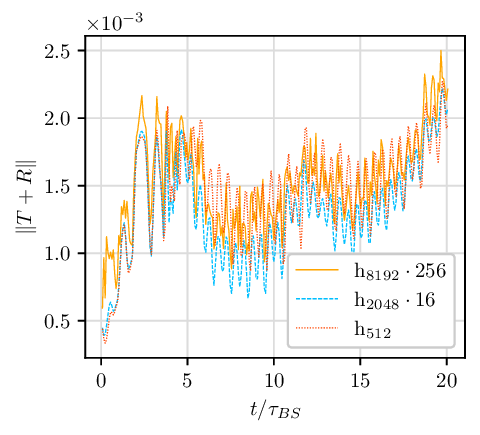}
\caption{Convergence of the $l_2$-norm of independent residuals for the trace 
of the Einstein field equations ($T+R = 0$) in the case of a very strongly 
perturbed d-star. The residuals of the higher resolution simulations are 
scaled by 16 and 256, respectively such that overlap of the curves implies 
second order convergence. This figure encompasses approximately 50 
light-crossing times or 20 periods of the central boson star oscillation, 
$\tau_{\mathrm{BS}}$.}
\label{bsgm_l_convergence_ire}
\end{figure}

\begin{figure}[!htb]
\centering
\includegraphics[scale=1.0]
{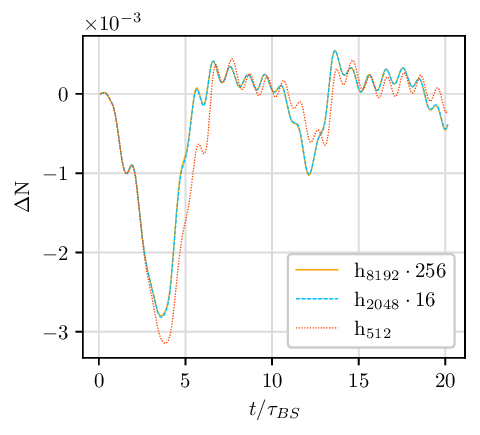}
\caption{Convergence of charge conservation for strong field data where the 
residuals of the higher resolution simulations have been scaled such that 
overlap of the curves implies second order convergence. The oscillation 
period of the unperturbed central boson star is denoted 
$\tau_{\mathrm{BS}}$.  }
\label{bsgm_l_conservation_noether}
\end{figure}

\begin{figure}[!htb]
\centering
\includegraphics[scale=1.0]
{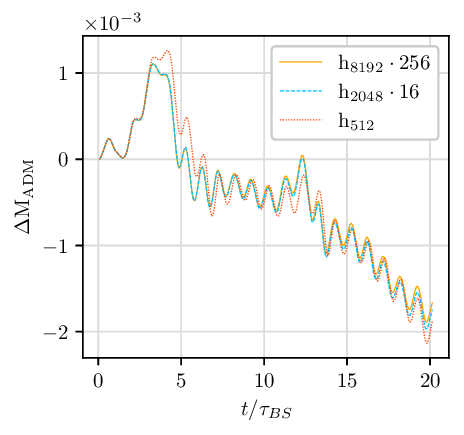}
\caption{Convergence of mass conservation for strong field data  where the 
residuals of the higher resolution simulations have been scaled such that 
overlap of the curves implies second order convergence.  }
\label{bsgm_l_conservation_mass}
\end{figure}

We demonstrate the convergence of our algorithm via the evolution of slightly 
sub-critical (i.e.~slightly stronger initial perturbations would result in 
black hole formation), non-minimally coupled, initial data. In verifying the 
validity of our evolutionary scheme, we make use of the technique of 
independent residual evaluators. This involves creating alternative 
discretizations of the equations of motion (EOM) which are then applied to 
solutions computed via our evolutionary scheme. More explicitly, our 
evolutionary scheme solves the difference equations,
\begin{align}
   \tilde{D}(u^h) - f^h = 0,
\end{align}
where $\tilde{D}$ is some non-linear difference operator and $u^h$ and $f^h$ 
are our discretized fields with grid spacing $h$. The technique of 
independent residual evaluation involves finding an alternative 
discretization, $\tilde{D}'$, application of which to our pre-computed 
solution, $u^h$, yields,
\begin{align}
   \tilde{D}'(u^h) - f^h = \tilde{r}^h.
\end{align}
If $\tilde{r}^h$ is observed to converge at order $O(h^2)$, it implies that 
both $\tilde{D}$ and $\tilde{D}'$ match to order $O(h^2)$ and provides 
confidence---far beyond what can be achieved with standard convergence 
tests---that we are solving the correct EOM. 
Figure~\ref{bsgm_l_convergence_ire} shows representative independent residual 
convergence of strong field initial data for a range of grid spacings while 
Figs.~\ref{bsgm_l_conservation_noether} and \ref{bsgm_l_conservation_mass} 
plot conserved quantity violations for the same solutions. From these plots, 
it can be seen that the solution algorithm is convergent in the strong field 
limit.

\subsection{Extraction of Growth Modes\label{bsgm_l_subsec_ds_growth_modes}}

The basic mechanics of a perturbation theory analysis  suggest an obvious means 
by which the stability of a solution may be tested. By monitoring the growth 
rate of a quantity which would remain constant were the solution unperturbed, 
we can make a direct measurement of the eigenvalue of the dominant mode. 
In the case of instability, we expect to see exponential growth in the norm 
of the perturbation. In contrast, the norm of a perturbed quantity which
is stable should oscillate in time. In both cases, the growth rate or period, 
respectively, may be determined by appropriate fits to the perturbed quantity.

A note of caution is, however, warranted: the method described above is only 
capable of determining a {\em lower} bound for the instability of the 
system. It is entirely possible for this method to miss unstable modes with 
eigenvalues much smaller than those of the excited stable modes (in which 
case the growth rate of the former is masked by oscillations of the latter). 
In an effort to counter this problem, we evolve the perturbed solutions for 
many dynamical timescales as given by the lowest frequency stable mode. This 
typically translates to hundreds of light-crossing times and thousands of 
boson star oscillations. Although we can never state with absolutely 
certainty that a solution is stable, we will nonetheless use that terminology 
when no growing modes are detected over such time scales.

Previous stability studies~\cite{lee1989stability, kleihaus2012stable, 
gleiser1989gravitational, kusmartsev1991gravitational} have shown that boson 
stars, like fluid stars, undergo stability transitions only at solutions 
corresponding to extrema of the asymptotic mass, $M_\infty$, as is predicted 
by catastrophe theory. We therefore work under the assumption that all sets 
of stationary solutions in the same region (bounded by extrema of $M_\infty$) 
exhibit similar stability properties. The results of 
Sec.~\ref{bsgm_l_subsec_results_lpt} serve to validate this assumption for the 
minimally coupled case.
The results presented in Sec.~\ref{bsgm_l_subsec_results_ds} are therefore 
derived from a small number of  simulations chosen to be representative of 
each region in a given family. Typically, we perform between two and three 
simulations for each region.

In our analysis we make extensive use of the Noether charge as a stability 
diagnostic rather than using matter fields or metric functions. 
Given~(\ref{bsgm_l_noether}), we see that the Noether charge is a derived 
quantity, tied tightly to both the metric functions and bosonic matter fields. 
As such, any changes to those fields are immediately reflected in the Noether 
charge, making it an ideal quantity for monitoring stability and for comparing 
the form of excited modes to those predicted by perturbation theory.

\section{Linear Perturbation Theory\label{bsgm_l_sec_lpt}}

Proceeding in standard fashion, we decompose the perturbed solution, 
$f(t,r)$, into a stationary component and an integral over Fourier modes:  
\begin{align}
   f(t,r)=f_0(r)+\int_{-\infty}^{\infty}\hat{f}(r, \beta)e^{i\beta t} d\beta.
\end{align}
We substitute $f(t,r) = f_0(r)+ \epsilon \delta f(r)e^{i\beta t}$ into the 
equations of motion and expand to linear order order in $\epsilon$. By doing 
so, we reduce our system of PDEs to a system of ODEs 
that represents the growth 
rate of various modes and that constitutes  an eigenvalue problem in $\beta^2$. 
In general, this system cannot be solved for all values of $\beta^2$ while 
retaining conservation of our conserved quantities; for most values of 
$\beta^2$, the solution obtained implies that the integral of the various 
conserved quantities is time dependent. For those countable number of modes 
that do satisfy the requisite boundary conditions for the conserved quantities, 
those with $\beta^2>0$ will be stable while those with $\beta^2 < 0$ will 
be unstable.

Following Gleiser and Watkins~\cite{gleiser1989gravitational}, we transform
to a set of variables $(\mu(t,r)$, $\nu(t,r)$, $\psi_R(t,r)$, $\psi_I(t,r)$, 
$\phi(t,r))$ defined by:
\begin{align}
   a &= e^{\mu/2},
\\
   \alpha &= e^{\nu/2},
\\
   \Psi &= e^{-i\omega t}\left(\psi_R+i\psi_I\right),
\\
   \phi_M &= \phi.
\end{align}
The derivation and final form of the perturbation equations have been 
relegated to Appendix \ref{bsgm_l_app_linear_perturbation_theory_equations} 
due to their significant complexity.

\subsection{Solution Procedure\label{bsgm_l_subsec_lpt_solution_procedure}}

Even in the minimally coupled case, finding solutions 
to~(\ref{bsgm_l_app_pert_start})--(\ref{bsgm_l_app_pert_end}) proves to be 
quite challenging. Examination of the regularity conditions at the origin 
reveals that the appropriate degrees of freedom in the problem are given 
by $\beta^2$, $\delta\psi_R(r)|_{r=0}$, $\delta\psi_I(r)|_{r=0}$ and 
$\p_r{\delta\phi(r)|_{r=0}}$. Due to the linearity of the problem, we are 
free to set $\delta\psi_R(r)|_{r=0} = 1$ and the equations therefore 
constitute 
an eigenvalue-boundary value problem in the remaining degrees of freedom.
In general, for inexactly chosen boundary values and $\beta^2$, the matter 
and metric 
functions may only be integrated to a finite distance from the origin 
before the solution becomes pathological. As such, we cannot use, for example, 
gradient descent techniques to tune these parameters and instead turn 
to an iterative shooting method~\cite{gdreid_2016}.

We expect that each successive mode of solutions 
to~(\ref{bsgm_l_app_pert_start})--(\ref{bsgm_l_app_pert_end}) will 
develop an additional node in each of the field variables. Correspondingly, 
we choose a trial value of $\beta^2$ (manually due to the difficulty 
encountered automating the process) and set $\delta\phi(r)=0$. Once the fields 
have been initialized, we shoot on $\delta\psi_I(r)|_{r=0}$ for $\delta\psi_R$ 
and $\delta\psi_I$ holding $\delta\phi$ fixed until the bosonic fields are well 
behaved far from the origin. We then fit a decaying tail to $\delta\psi_R$ and 
$\delta\psi_I$, and shoot on 
$\p_r{\lp(\delta\phi\rp)}|_{r=0}$ holding the bosonic fields fixed. This 
shooting procedure is repeated several times until approximate convergence is 
achieved. Finally, we examine $\delta N_\infty$ and adjust $\beta^2$, 
repeating the entire shooting procedure until a solution is found with 
$\delta N_\infty \approx 0$ to within tolerance (typically 0.05 of the 
maximum value of $\delta N(r)$ is sufficient). This process of determining 
initial data in compactified coordinates $x$ is summarized in 
Algorithm~\ref{bsgm_l_iterated_shooting_algorithm}.

\begin{figure}[!ht]
\begin{algorithm}[H]
\caption{Iterated Shooting Procedure}\label{bsgm_l_iterated_shooting_algorithm}
\begin{algorithmic}
\State {\textbf{initialize} background fields on the compactified grid $x$}
\State {\textbf{initialize} $\delta\phi(x)$ to 0}
\State {\textbf{initialize} $\delta\psi_R(x)$ and $\delta\psi_I(x)$ to 0}
\While{$\delta N_{\infty} \gg 0$}
\State{\textbf{choose} $\beta^2$}
\While{solution non-convergent}
\State {\textbf{hold} $\delta\phi(x)$ fixed}
\State {\textbf{shoot} for $\delta\psi_R(x)$ and $\delta\psi_I(x)$}
\State {\textbf{fit} tail to $\delta\psi_R(x)$ and $\delta\psi_I(x)$}
\State {\textbf{integrate} $\delta\nu(x),\delta\mu(x)$ and $\delta N$ to}
\Statex{\quad\quad\quad asymptotic region}
\Statex{}
\State {\textbf{hold} $\delta\psi_R(x)$ and $\delta\psi_I$ fixed}
\State {\textbf{shoot} for $\delta\phi(x)$}
\State {\textbf{fit} tail to $\delta\phi(x)$}
\State {\textbf{integrate} $\delta\nu(x),\delta\mu(x)$ and $\delta N$ to}
\Statex{\quad\quad\quad asymptotic region}
\EndWhile
\EndWhile
\end{algorithmic}
\end{algorithm}
\end{figure}

Upon achieving the desired tolerance, the approximate solution is used as 
an initial guess for a boundary value problem solver based on the collocation 
library TWPBVPC~\cite{MazziaTWPBVPC}. If the initial guess is sufficiently 
close to the true solution, the solver converges quickly, resulting in a 
solution which is accurate to within tolerance (typically $10^{-12}$ or 
better). 
By slowly adjusting the central amplitude, $\psi(0)$, of the stationary 
solution and using the previous solution to the perturbative equations as an 
initial guess, we can use the process of continuation to investigate the 
development of the mode throughout a branch of a family. \footnote{From a 
technical perspective, it is worth noting how the $\beta^2$ eigenvalue is 
incorporated in {TWPBVPC}. Following~\cite{anja_email}, we implemented the 
$\beta^2$ eigenvalue as an additional field satisfying the trivial equation 
$\partial_x(\beta^2(x))=0$.}

As discussed in Sec.~\ref{bsgm_l_sec_ds}, we only perform 2 or 3 evolutions 
per region to assess dynamical stability.  In contrast, the process of 
continuation gives a much more comprehensive view of the mode structure 
within a region.
By repeating this procedure on every branch and near every extremal point of 
$M_\infty$ (including discontinuities), we can achieve an accurate picture 
of the mode structure of the family under investigation.

Note also that, due to the presence of asymptotic shells of matter which are 
present in some of these solutions, traditional shooting techniques may fail 
to adequately resolve the perturbed boson fields. The inability of double 
precision shooting to provide an adequate initial guess to the {TWPBVPC} based 
solver may be resolved, to some degree, through the use of extended precision 
integrators or through the use of our multi-precision shooting 
method~\cite{gdreid_2016}. Even then, we find that perturbative solutions in 
the presence of an asymptotic shell are quite difficult to find without the 
aid of continuation.

\subsection{Convergence\label{bsgm_l_subsec_lpt_convergence}}

\begin{figure}[!htb]
\centering
\includegraphics[scale=1.0]
{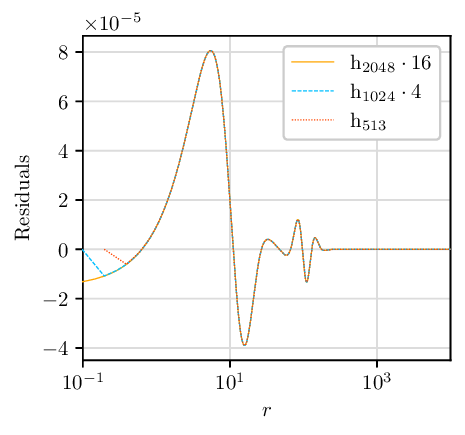}
\caption{Convergence of independent residuals for $\delta \lambda$ from 
family $p_1$ for a stable mode corresponding to $\psi(0) = 0.030$ and 
$\beta^2 \approx 3.21\cdot10^{-6}$. Here we plot the scaled residuals of the 
metric function evaluated on grids of 2049, 1025 and 513 points using a 
second order finite difference scheme for the independent residual evaluator. 
With the scaling given in the figure, overlap of the curves implies 
second order convergence.}
\label{bsgm_l_perturbation_convergence}
\end{figure}

As with the dynamical evolutions, we verify the convergence of our 
perturbative solutions via an independent residual convergence test. In this 
case, as in~\cite{gdreid_2016}, care must be taken when computing these 
residuals due to the method by which TWPBVPC determines solutions. By 
default, TWPBVPC attempts to minimize solution error with a deferred error 
correction scheme that uses a combination of high order discretizations and 
allocation of additional grid points in the vicinity of poorly resolved 
features. Although these properties are invaluable for producing high quality 
solutions, they serve to increase the effective resolution and convergence 
order of a solution, making independent residual convergence difficult to 
verify. Consequently, Fig.~\ref{bsgm_l_perturbation_convergence} demonstrates 
the convergence of our collocation code with deferred error correction 
disabled.

\section{Results\label{bsgm_l_results}}

Section~\ref{bsgm_l_subsec_results_ds} presents the results of our dynamic 
simulations and summarizes the regions of stability found for the families of 
Table~\ref{bsgm_l_table_families}. Section.~\ref{bsgm_l_subsec_results_lpt} 
summarizes the results of our perturbation theory analysis for families $p_1$ 
and $p_2$. We derive a more complete picture of the modal structure for these 
families and provide important insight into the 
stability of other families of initial data. Finally, 
Sec.~\ref{bsgm_l_subsec_results_compare} compares the results from dynamical 
simulations and perturbation theory for family $p_1$, demonstrating the broad 
equivalence of the two methods.

\subsection{Dynamical Simulations\label{bsgm_l_subsec_results_ds}}

Using slightly perturbed stationary solutions as initial data, 2 or 3 long 
time simulations for each region of each family in 
Table~\ref{bsgm_l_table_families} were performed and the growth of 
perturbations in $N$, $\Psi\Psi^*$, $\Phi_M$, $a$ and $\alpha$  were 
monitored. Simulations exhibiting collapse, dispersal or 
non-stationary remnants were deemed unstable. Conversely, those showing 
oscillation about the stationary solutions with an oscillation magnitude set 
by the size of the initial perturbation were deemed stable.
Again, we assume that the stability properties of all configurations in a 
given region are the same but perform a minimal validation of this assumption 
by performing at least 2-3 simulations per region.

In the interest of minimizing errors originating from our use of a dissipative 
second order code, these representative simulations were performed only near
the centers of regions, fairly distant in parameter space from turning 
points of the mass and from branch jumps. In the case of regions with 
asymptotic shells, we concentrated our simulation efforts on areas where the 
d-stars were reasonably compact. In doing so, it was possible to uniformly 
excite modes and ensure that the light-crossing times for the compact objects 
were much less than the simulation time.

\begin{figure}[!ht]
\centering
\includegraphics[scale=1.0]
{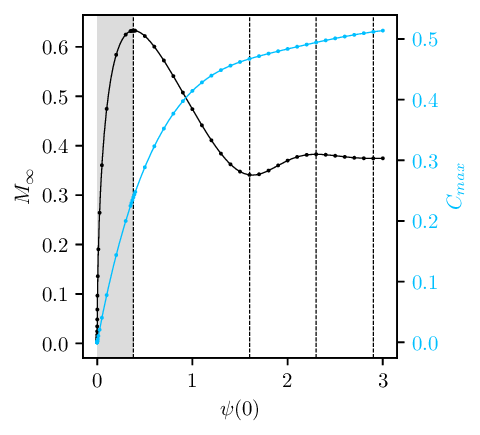}
\caption{Combined asymptotic mass and maximum compactness plot for the family 
of mini-boson stars (minimally coupled boson stars in the absence of a global 
monopole). The region of stability is denoted in gray. Dashed lines show 
turning points of the mass.}
\label{bsgm_l_stability_plot_boson}
\end{figure}

\begin{figure}[!ht]
\centering
\includegraphics[scale=1.0]
{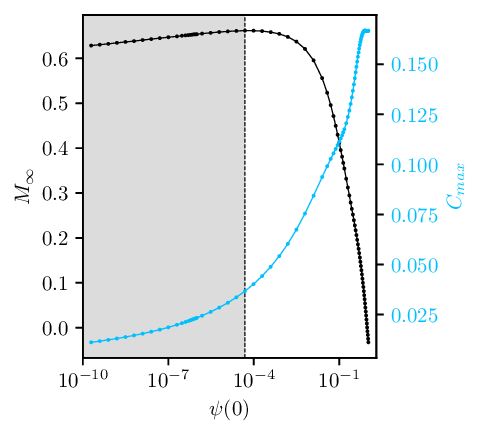}
\caption{Combined asymptotic mass and maximum compactness plot for family 
$c$ ($\Delta^2 = 0.36$, $\lambda_{G} = 1.000$, $\xi_{B} = 0$, $\xi_{G} = 0$). 
The region of stability is denoted in gray. Dashed lines show turning points 
of the mass.}
\label{bsgm_l_stability_plot_4}
\end{figure}

\begin{figure}[!ht]
\centering
\includegraphics[scale=1.0]
{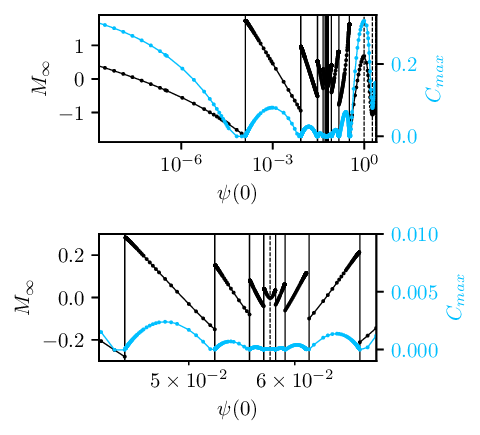}
\caption{Combined asymptotic mass and maximum compactness plot for family 
$d$ ($\Delta^2 = 0.81$, $\lambda_{G} = 0.010$, $\xi_{B} = 0$, $\xi_{G} = 0$). 
No region of stability is found. Dashed vertical lines show turning points of 
the mass while solid vertical lines denote boundaries of solution branches.
Here and in the next two plots the bottom panel shows a zoomed-in view of 
a portion of the data in the top panel.
   }
\label{bsgm_l_stability_plot_5}
\end{figure}

\begin{figure}[!ht]
\centering
\includegraphics[scale=1.0]
{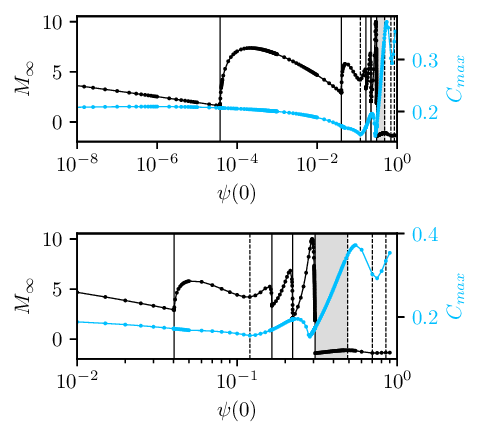}
\caption{Combined asymptotic mass and maximum compactness plot for family 
$e$ ($\Delta^2 = 0.25$, $\lambda_{G} = 0.001$, $\xi_{B} = 3$, $\xi_{G} = 3$). 
The region of stability is shown in gray. Dashed vertical lines show turning 
points of the mass while solid vertical lines denote boundaries of solution 
branches.}
\label{bsgm_l_stability_plot_6}
\end{figure}

\begin{figure}[!ht]
\centering
\includegraphics[scale=1.0]
{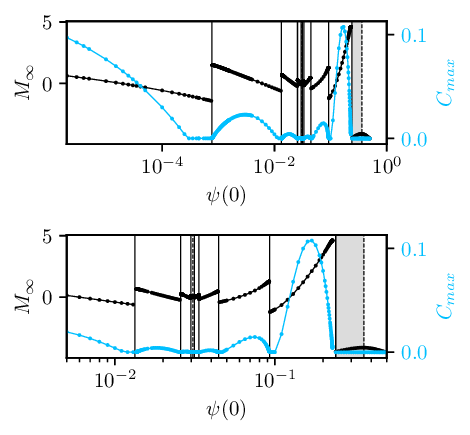}
\caption{Combined asymptotic mass and maximum compactness plot for family 
$f$ ($\Delta^2 = 0.49$, $\lambda_{G} = 0.010$, $\xi_{B} = 5$, $\xi_{G} = 0$). 
The region of stability is shown in gray. Dashed vertical lines show turning 
points of the mass while solid vertical lines denote boundaries of solution 
branches.}
\label{bsgm_l_stability_plot_7}
\end{figure}

\begin{figure}[!ht]
\centering
\includegraphics[scale=1.0]
{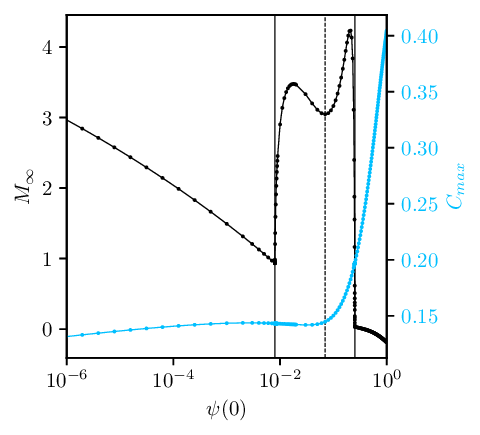}
\caption{Combined asymptotic mass and maximum compactness plot for family 
$g$ ($\Delta^2 = 0.09$, $\lambda_{G} = 0.010$, $\xi_{B} = 0$, $\xi_{G} = 5$). 
No region of stability is found. Dashed vertical lines show turning points of 
the mass while solid vertical lines denote boundaries of solution branches.}
\label{bsgm_l_stability_plot_8}
\end{figure}

\begin{figure}[!ht]
\centering
\includegraphics[scale=1.0]
{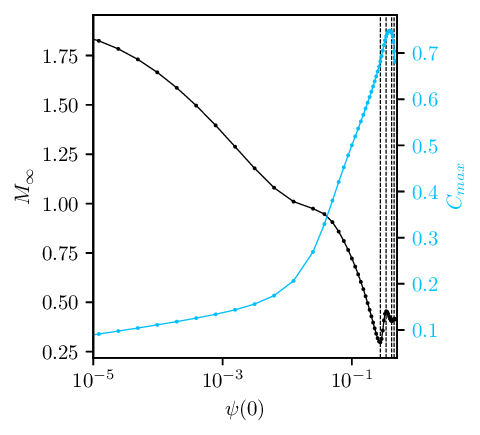}
\caption{Combined asymptotic mass and maximum compactness plot for family 
$h$ ($\Delta^2 = 0.08$, $\lambda_{G} = 0.10$, $\xi_{B} = -4$, $\xi_{G} = 5$). 
No region of stability is found. Dashed vertical lines show turning points of 
the mass.}
\label{bsgm_l_stability_plot_15}
\end{figure}

\begin{figure}[!ht]
\centering
\includegraphics[scale=1.0]
{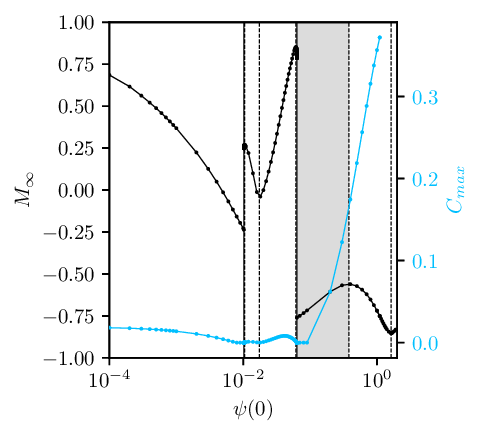}
\caption{Combined asymptotic mass and maximum compactness plot for family 
$p_1$ ($\Delta^2 = 0.09$, $\lambda_{G} = 0.04$, $\xi_{B} = 0$, $\xi_{G} = 0$). 
The region of stability is shown in gray. Dashed vertical lines show turning 
points of the mass while solid vertical lines denote boundaries of solution 
branches.}
\label{bsgm_l_stability_plot_p1}
\end{figure}

\begin{figure}[!ht]
\centering
\includegraphics[scale=1.0]
{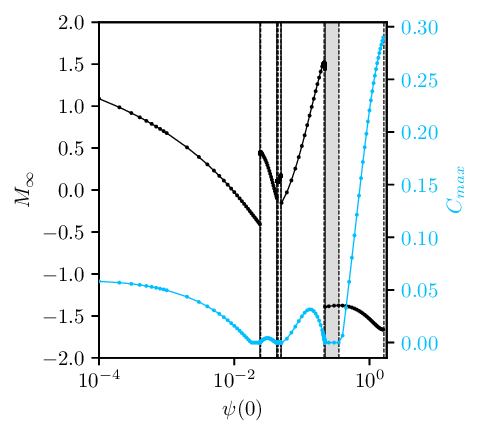}
\caption{Combined asymptotic mass and maximum compactness plot for family 
$p_2$ ($\Delta^2 = 0.25$, $\lambda_{G} = 0.04$, $\xi_{B} = 0$, $\xi_{G} = 0$). 
The region of stability is shown in gray. Dashed vertical lines show turning 
points of the mass while solid vertical lines denote boundaries of solution 
branches.}
\label{bsgm_l_stability_plot_p2}
\end{figure}

Figures~\ref{bsgm_l_stability_plot_boson}--\ref{bsgm_l_stability_plot_p2} are 
composite plots showing both the asymptotic mass, $M_{\infty}$, and maximum 
compactness, $C_{\mathrm{max}}$ as a function of boson star central amplitude
for the families listed in Table~\ref{bsgm_l_table_families}. Regions 
highlighted in gray are stable under small radial perturbations while regions 
outside the gray shading are unstable. 
The data points plotted here are drawn from our calculations of stationary 
solutions, not the much more sparsely sampled dynamical simulations. As noted 
above, the stability of each region was determined using far fewer 
simulations than there are points on the graph.

Special attention should be paid to Fig.~\ref{bsgm_l_stability_plot_boson} 
which shows the asymptotic mass, maximum compactness and regions of 
stability for the case of the mini-boson star. For mini-boson stars 
themselves, it is the region before the first turning point in the  mass 
that is stable~\cite{gleiser1989gravitational, kusmartsev1991gravitational}. 
In all d-stars investigated, the stable region, when it exists, corresponds to 
the region immediately before the first turning point on the 
final branch (where $\psi(0)$ assumes its largest values). Since we have 
previously shown that this final branch has no shells of bosonic matter far 
from the origin~\cite{gdreid_2016}, we find that, for  all families of 
d-stars so far investigated, regions of stability are confined to 
boson-star-like branches without asymptotic shells.

\subsection{Linear Perturbation Theory\label{bsgm_l_subsec_results_lpt}}

In performing our perturbation theory analysis, we have restricted our 
investigation of d-stars to two minimally coupled families of solutions 
designated as $p_1$ and $p_2$ in Table~\ref{bsgm_l_table_families}. These 
families were chosen for two primary reasons. First, they have relatively 
simple branching structures (shown in Figs.~\ref{bsgm_l_stability_plot_p1} 
and \ref{bsgm_l_stability_plot_p2}) and this simplifies the perturbation 
analysis. Second, the two families are very close to one another in parameter 
space, yet have different numbers of solution branches. Correspondingly, 
their analysis yields clues as to how the modal structure changes as we 
vary parameters other than the family parameter, $\psi(0)$. For comparison 
purposes, we also include the results of perturbation theory applied to the 
case of minimally coupled mini-boson stars.

\begin{figure}[!htb]
\centering
\includegraphics[scale=1.0]
{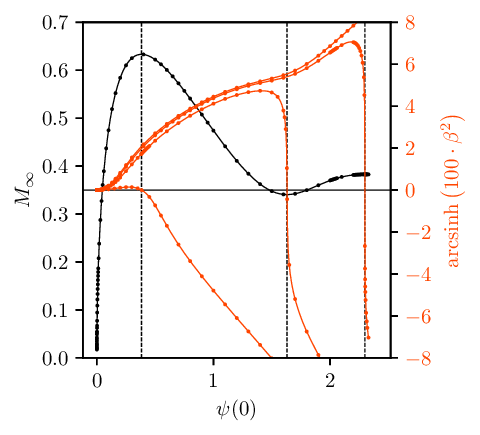}
\caption{
Eigenvalues for the family of mini-boson stars. The modal structure shown here 
contrasts with behaviour presented in 
Figs.~\ref{bsgm_l_beta2_family1}--\ref{bsgm_l_beta2_family2} for minimally 
coupled d-stars. Note that at each turning point of the mass (vertical black 
dashed lines), a stable mode transitions to unstable. Here, and in subsequent 
plots, eigenvalues, $\beta^2$, are shown as functions of 
$\sinh{(\lambda \beta^2)}$ to better display the overall modal structure: the 
magnitude of $\beta^2$ varies greatly so $\lambda$ is chosen on a 
plot-by-plot basis to more clearly show the overall behaviour of the 
eigenvalues.
}
\label{bsgm_l_beta2_boson}
\end{figure}

\begin{figure}[!htb]
\centering
\includegraphics[scale=1.0]
{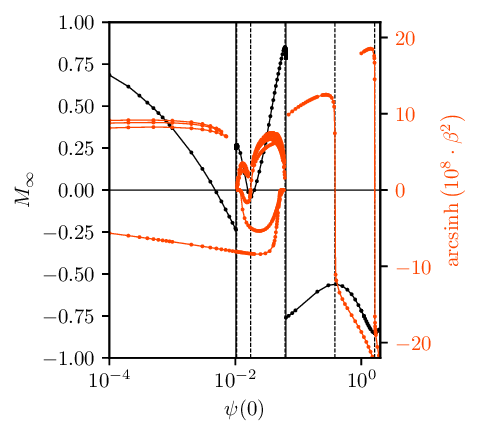}
\caption{Eigenvalues of family $p_1$ as a function of the boson star central 
amplitude, $\psi(0)$. Note the apparent discontinuities in the eigenvalues 
near branch transitions. In these regions, the eigenvalues become
near-degenerate and our solutions are no longer convergent. Our observations, 
however, are consistent with the stable eigenvalues approaching $\beta^2=0$ 
at the branch transitions. Note also that there are an infinite number of 
stable modes in each region; here we have plotted only the first three. }
\label{bsgm_l_beta2_family1}
\end{figure}

\begin{figure}[!htb]
\centering
\includegraphics[scale=1.0]
{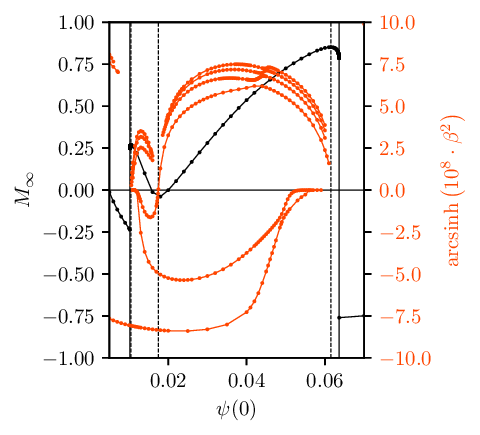}
\caption{A more detailed view of the central region of 
Fig.~\ref{bsgm_l_beta2_family1} for family $p_1$. Interestingly, the central 
branch is composed of four regions (rather than two), with the first and last 
corresponding to shells very far from the origin. Unfortunately, the 
eigenvalue degeneracy prevents us from investigating these regions in detail, 
but it is regardless evident that the unstable modes persist through these 
regions (with the potential exception of the very last region where we were 
unable to resolve any perturbations). }
\label{bsgm_l_beta2_family1_zoom}
\end{figure}

\begin{figure}[!htb]
\centering
\includegraphics[scale=1.0]
{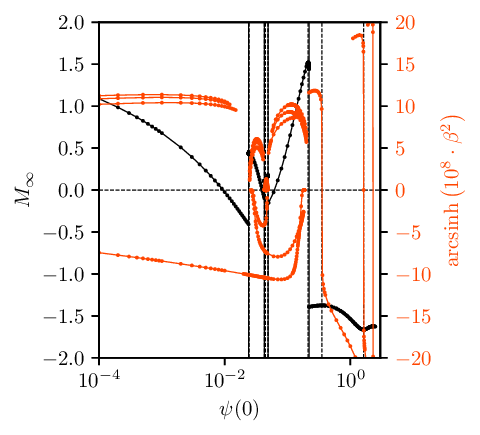}
\caption{Eigenvalues of family $p_2$ as a function of the boson star central 
amplitude, $\psi(0)$. The  additional two branch transitions have split the 
second branch of family $p_1$ into three distinct regions (see 
Fig.~\ref{bsgm_l_beta2_family1}). Note that the region after the final branch 
transition is qualitatively very similar to that of mini-boson stars as 
shown in Fig.~\ref{bsgm_l_beta2_boson}. As for family $p_1$, it can be seen 
that the central branches exhibit turning points in the mass corresponding 
to shells very far from the origin.}
\label{bsgm_l_beta2_family2}
\end{figure}

\begin{figure}[!htb]
\centering
\includegraphics[scale=1.0]
{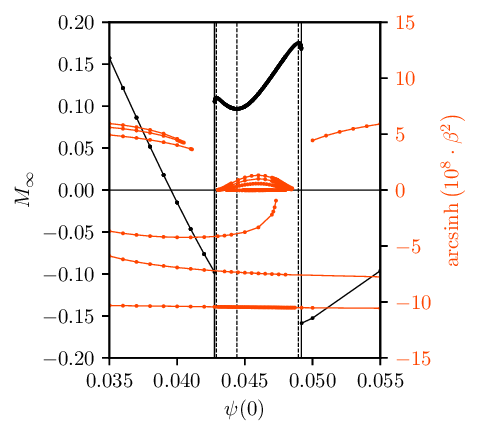}
\caption{Expanded view of Fig.~\ref{bsgm_l_beta2_family2} showcasing details 
that were poorly resolved in the original plot. Our simulation data is 
consistent with the eigenvalues of the poorly resolved modes  approaching 0. 
As for family $p_1$, it can be seen that the central branches exhibit turning 
points in the mass corresponding to shells very far from the origin. 
Unfortunately, the extreme length scales in these solutions, coupled with the 
eigenvalue degeneracy of the stable modes, prevents us from examining these 
regions in more detail. }
\label{bsgm_l_beta2_family2_zoom}
\end{figure}

Plots displaying the modal structure for families $p_1$ and $p_2$ as well as 
for mini-boson stars are shown in 
Figs.~\ref{bsgm_l_beta2_boson}--\ref{bsgm_l_beta2_family2_zoom}, which plot
eigenvalues, $\beta^2$, as a function of the boson star central amplitude, 
$\psi(0)$. Stable regions have only modes with $\beta^2 > 0$ while unstable 
regions have modes both with $\beta^2 > 0$ and $\beta^2 < 0$. Note that the 
complete spectral structures are not shown. Rather, only the first few 
modes (the least stable) are displayed in each case.

In Figs.~\ref{bsgm_l_beta2_family1}--\ref{bsgm_l_beta2_family2_zoom} we 
follow the convention of the previous sections and display the locations of 
branch transitions (corresponding to discontinuities in the mass and Noether 
charge) as black vertical lines while extrema of the mass are given as black 
dashed vertical lines. We observe that stability transitions within a 
branch occur  at these extremal points as predicted by catastrophe 
theory~\cite{kusmartsev1991gravitational, kleihaus2012stable}. 

Near the branch transitions, the eigenvalues of many modes appear to become 
degenerate, and neither our collocation or evolutionary codes are capable of 
investigating these regions in much detail. What is obvious, however, is that 
there is a well resolved unstable mode which persists across all but the final 
branch of families $p_1$ and $p_2$. The presence of this unstable mode 
indicates that stable solutions exist only on the final branch, in perfect 
agreement with our dynamical simulations.

The lack of resolution resulting from the eigenvalue degeneracies in the 
vicinity of the branch transitions somewhat complicates the interpretation. 
We thus present a heuristic argument to build up a generic picture of the mode 
transitions. Recall from~\cite{gdreid_2016} that each branch transition 
before the mass turning point corresponds to the formation of a bosonic shell 
at infinity. This shell then migrates inward as the family parameter, 
$\psi(0)$, is increased. Each branch transition after the mass extrema then 
corresponds to the disappearance of a bosonic shell at infinity after it 
migrates outwards.  

\begin{figure}[!ht]
\centering
\includegraphics[scale=1.0]
{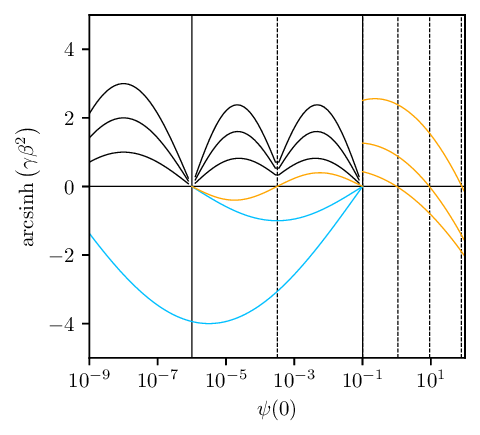}
\caption{Idealized plot of eigenvalues highlighting the underlying structure 
of family $p_1$. Here we plot stable modes in black, unstable modes in blue 
and modes which undergo a stability transition within a branch in orange. Note 
in particular how in a family ``close'' to developing an additional branch 
transition, the eigenvalues ``near'' where the transition would develop become 
increasingly degenerate. }
\label{bsgm_l_beta2_example1}
\end{figure}

In the case of families $p_1$ and $p_2$, the first and final transitions can 
be identified with the appearance and disappearance of the first shell of 
matter. For family $p_2$, the second and third transitions can likewise be 
identified with the appearance and disappearance of a second shell of matter. 
As noted above, family $p_1$ is very close to developing additional branch 
transitions similarly to family $p_2$. For a marginally larger value of 
$\Delta^2$, family $p_1$ would have a degenerate transition corresponding to 
a shell of matter which appears suddenly at infinity and then immediately 
vanishes as the family parameter, $\psi(0)$, is increased. We can, in fact, 
see evidence of this behaviour in Fig.~\ref{bsgm_l_beta2_family1} and 
idealized in Fig.~\ref{bsgm_l_beta2_example1} where the eigenvalues of the 
stable modes dip down towards 0 near the mass extrema.

\clearpage

\begin{figure}[!ht]
\centering
\includegraphics[scale=1.0]
{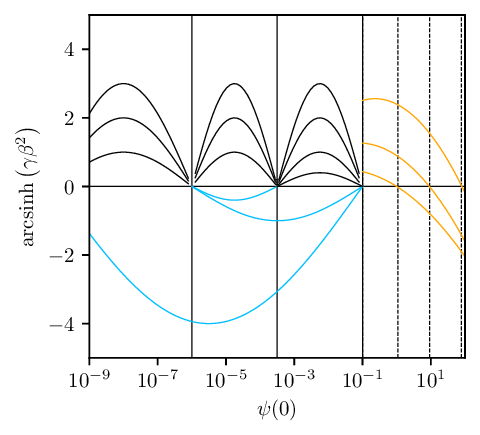}
\caption{Idealized plot of eigenvalues in the case of a degenerate branch 
transition. The transitioning mode has been split into a stable and unstable 
branch. As before, we plot stable modes in black, unstable modes in blue 
and modes which undergo a stability transition within a branch in orange.}
\label{bsgm_l_beta2_example2}
\end{figure}

\begin{figure}[!ht]
\centering
\includegraphics[scale=1.0]
{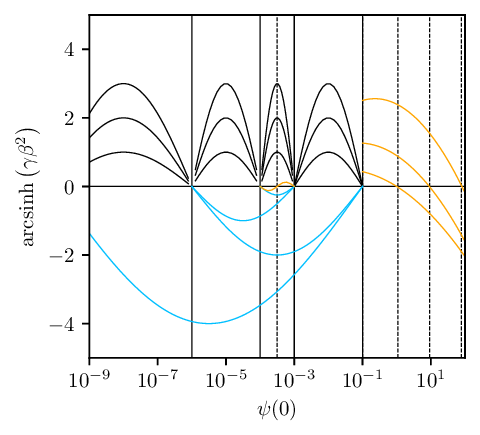}
\caption{Idealized plot of eigenvalues highlighting the underlying structure 
of family $p_2$. As $\Delta^2$ is increased further relative to $p_1$, the 
degeneracy of Fig.~\ref{bsgm_l_beta2_example2} is resolved, and we gain a 
new branch of stable and unstable modes corresponding to the new shell of 
matter.}
\label{bsgm_l_beta2_example3}
\end{figure}

This degenerate transition would split the transitioning mode into a stable 
and unstable region as shown in Fig.~\ref{bsgm_l_beta2_example2}. As 
$\Delta^2$ is increased further, the degeneracy is resolved, and we gain a 
new unstable mode corresponding to the new shell of matter as shown in 
Fig.~\ref{bsgm_l_beta2_example3}. Given that there is still a mass extrema 
between the branch transitions, we will additionally have a new 
transitioning mode which changes from stable to unstable in accordance 
to catastrophe theory~\cite{kusmartsev1991gravitational, kleihaus2012stable}. 
The final picture is then a series of stable eigenvalues between each branch 
transition joined via unstable modes in the manner depicted in 
Fig.~\ref{bsgm_l_beta2_example3}. 

\subsection{Comparison of Methods\label{bsgm_l_subsec_results_compare}}

\begin{figure}[!ht]
\centering
\includegraphics[scale=1.0]
{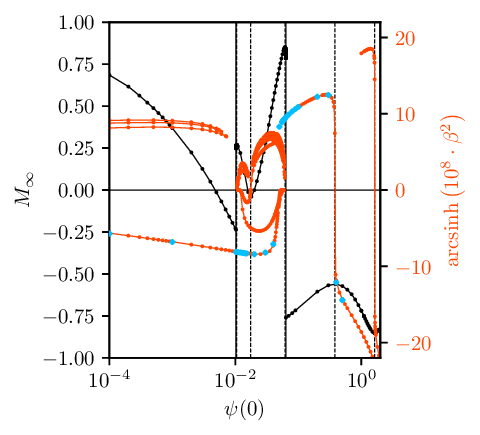}
\caption{Mass and mode structure for family $p_1$. The asymptotic mass and 
eigenvalues measured via perturbation theory are shown in black and red as 
in Fig.~\ref{bsgm_l_beta2_family1}. Eigenvalues measured directly from 
dynamical simulation are shown as blue diamonds. In the second branch, the 
blue diamonds indicating the existence of an oscillatory mode with 
$\beta^2 > 0$ do not correspond to any perturbative mode that we were able 
to identify. }
\label{bsgm_l_beta2_family1_combine}
\end{figure}

Using the method of Sec.~\ref{bsgm_l_subsec_results_ds}, we extract the 
eigenvalues of the most unstable modes of the dynamical simulations for family 
$p_1$ and plot them against the results predicted by perturbation theory in the 
previous section. In Fig.~\ref{bsgm_l_beta2_family1_combine}, eigenvalues
from individual dynamical simulations are shown as blue circles while the 
eigenvalues from perturbation theory are shown in red. 

Examining Fig.~\ref{bsgm_l_beta2_family1_combine}, we can see that the 
perturbation theory and the dynamical simulation are in agreement for the 
majority of the parameter space. An interesting issue, however, arises just 
before the final branch. 
In this region, the dynamical simulations suggest the existence of a stable 
oscillatory mode and the absence of the unstable mode found through 
perturbation theory. 

This discrepancy is likely the result of two confounding factors. The first 
is that, in the region under consideration, the growth rate of the modes 
is extremely small ($|\beta^2| \lesssim 1\cdot 10^{-8}$) so that the presence 
of an excited oscillatory mode could easily overwhelm the signal. If we were 
to integrate these solutions for a sufficiently long period of time, it seems 
likely that it would be possible to resolve the unstable eigenvalues. 
Unfortunately, the limited nature of our second order finite-difference based 
dynamical code makes maintaining temporal coherence for sufficiently long 
periods of time impractical.
The second factor is that the oscillatory modes observed in this region are 
not purely oscillatory and instead decay slightly with time. As shown in 
Appendix~\ref{bsgm_l_app_linear_perturbation_theory_equations}, however, the 
Hermitian character of this system requires that perturbations are either 
purely oscillatory or exponential in character. It seems likely that the 
oscillatory signals in these regions correspond to perturbations 
qualitatively close to those on the final branch which only nearly
satisfy~(\ref{bsgm_l_app_pert_start})--(\ref{bsgm_l_app_pert_end}).

In addition to comparing the eigenvalues $\beta^2$, it is possible to compare 
the profiles of the perturbations directly. Recall that for perturbations near 
the stationary solutions (for a given stationary field $f(t,r)$) we have
\begin{align}
   f(t,r) &= f_0(r) + \delta f(r)e^{-\beta t}, 
   \\
   \partial_t f &\propto \delta f.
\end{align} 
As such, we may find an approximation to the Noether charge perturbation by 
taking the time derivative of the Noether charge. Doing so for families 
$p_1$ and $p_2$, we find quite good agreement even for solutions with fairly 
distant bosonic shells.

\section{Summary\label{bsgm_l_sec_summary}}

We have addressed the question of boson d-star stability and suitability as 
black hole mimickers proposed by Marunovic and Murkovic 
in~\cite{black_hole_mimiker}. Through simulations of a diverse families of 
initial data, we have demonstrated that regions of stability, where they 
exist, are confined to the boson-star-like final branch of a given family in 
both the minimally coupled and non-minimally coupled case.

We have verified this result in the case of minimally coupled families through 
a fairly comprehensive mode analysis of two families of initial data ($p_1$ 
and $p_2$) which are close to one another in phase space. This analysis 
supports the results of our dynamical simulations with both perturbation 
theory and direct simulation being in broad agreement. The only exception 
to this is in regions where the magnitude of the unstable eigenvalues are 
small. Our evolutionary code is ill-suited to investigate these regions.

We observed how the number of branches and asymptotic shells change as the 
solid angle deficit, $\Delta^2$, is changed between these two families. 
By analysing the differences in mode structure between families $p_1$ and 
$p_2$ we propose a mechanism by which the mode structure changes in response 
to the appearance or disappearance of an asymptotic shell of bosonic matter.

Overall, our results are consistent with the interpretation that the highly 
compact solutions discovered in~\cite{black_hole_mimiker} are unstable. As a 
result, these solutions are likely poor candidates for astrophysically relevant 
compact objects. Finally, we observe that the novel solutions with shells of 
bosonic matter far from the origin discovered in~\cite{gdreid_2016} are 
likewise unstable.

\begin{acknowledgments}
This research was supported by the Natural Sciences and Engineering Research 
Council of Canada (NSERC).  
\end{acknowledgments}

\appendix

\section{Equations of Motion\label{bsgm_l_app_eom}}

We start with the 3+1 equations in polar-areal coordinates and take 
$T_{\mu\nu}$ to be the combination $T_{\mu\nu} = T^B_{\mu\nu} + T^G_{\mu\nu}$ 
as defined in (\ref{bsgm_l_T_uv_B})--(\ref{bsgm_l_T_uv_G}). In this gauge, 
the evolution of the metric function, $a$, is governed by
\begin{align}
   \label{bsgm_l_app_at_simp}
   \p_t a & = \frac{T_{tr} a r}{2},
\end{align}
the Hamiltonian constraint is
\begin{align}
   \label{bsgm_l_app_HC_simp}
   \frac{2 r \p_r a}{a} &= 1 -a^2 +\frac{a^2r^2T_{tt}}{\alpha^2},
\end{align}
and the slicing condition takes the form
\begin{align}
   \label{bsgm_l_app_SC_simp}
   2r \p_r\alpha &= \frac{2r\alpha\p_r a}{a} +\lp(\alpha T_{rr}
   - \frac{a^2 T_{tt}}{\alpha}\rp)r^2 
   \\ \nonumber
   & \hphantom{=}
   + 2\alpha\lp(a^2-1\rp).
\end{align}
Note that (\ref{bsgm_l_app_at_simp}) is redundant but provides a useful 
non-trivial consistency test for the system. In both the evolutionary and 
stationary cases, the Hamiltonian constraint and polar-areal slicing 
condition may be arranged to give explicit expressions for $\partial_r a$ and 
$\partial_r \alpha$, respectively. In this form, the Hamiltonian constraint 
is completely independent of $\alpha$, and the constraints may be 
independently integrated. Moreover, since the constraints are first order 
in $r$, no boundary value solver is needed. 

To simplify the resulting equations, we define the following quantities:
\begin{align}
   \label{bsgm_l_app_zeta}
   \zeta&=\frac{1}{1+\xi_{B}\phi_R^2+\xi_{B}\phi_I^2+\xi_{G}\Delta^2
   \phi_M^2},
\\
   \label{bsgm_l_app_Lambda_r}
   \Lambda_r &= \lp( \phi_R \Phi_R + \phi_I\Phi_I\rp)\xi_{B} 
   +\xi_{G} \D^2 \phi_M \Phi_M,
\\
   \label{bsgm_l_app_Lambda_t}
   \Lambda_t &= \lp( \phi_R \Pi_R + \phi_I\Pi_I\rp)\xi_{B} 
   +\xi_{G} \D^2 \phi_M \Pi_M,
\\
   \label{bsgm_l_app_delta}
   \delta &=2 + 2\Lambda_r r \zeta.
\end{align}
Upon substitution of (\ref{bsgm_l_app_zeta})--(\ref{bsgm_l_app_delta}) into 
(\ref{bsgm_l_app_at_simp})--(\ref{bsgm_l_app_SC_simp}), we find the following 
equations for $a$ and $\alpha$:
\begin{align}
        \label{bsgm_l_app_at_eqn}
        \partial_t a&
        =\frac{2\xi_{G}\Delta^2\zeta r \alpha\left(\phi_M\partial_r\Pi_M
        +\Pi_M\Phi_M\right)}{\delta}
        \\ \nonumber
        & \hphantom{=}
        +\frac{2\xi_{B}\zeta r \alpha\left(\Pi_R\Phi_R+\Pi_I\Phi_I+\phi_R
        \partial_r\Pi_R+\phi_I\partial_r\Pi_I\right)}{\delta}
        \\ \nonumber
        & \hphantom{=}
        -\frac{2 \zeta r\alpha\left(\xi_{G}\Delta^2\phi_M\Pi_M+\xi_{B}\left(
        \phi_R\Pi_R
        +\phi_I\Pi_I\right)\right)\partial_r a}{\delta a}
        \\ \nonumber
        & \hphantom{=}
        +\frac{\zeta r \alpha\left(\Pi_R\Phi_R+\Pi_I\Phi_I+\Delta^2\Pi_M
        \Phi_M+
        \Pi_P\Phi_P\right)}{\delta},
\end{align}
\begin{align}
        \label{bsgm_l_app_HC_eqn_time}
        \frac{4 \p_r a}{r a^3}&=
        -\frac{2}{r^2}
        +\frac{2}{r^2 a^2}
        -\lp[ -2V - \frac{2\D^2{\phi_M}^2}{r^2} 
        \right.
        \\ \nonumber
        &\mathopen{}\left. \hphantom{=}
        +\frac{1}{a^2}\lp( \vphantom{\frac{1}{1}} -\lp(
        \Phi_R^2 + \Phi_I^2 + \Pi_R^2 + \Pi_I^2 + \Phi_P^2 +\Pi_P^2 
        \rp.\rp.\rp.
        \\ \nonumber
        &\mathopen{}\lp.\lp.\lp. \hphantom{=}
        + \D^2\lp[\Phi_M^2 + \Pi_M^2\rp]  \rp) 
        -4 \lp(\phi_R \p_r \Phi_R + \phi_I 
        \p_r \Phi_I
        \rp.\rp.\rp.
        \\ \nonumber
        &\mathopen{}\lp.\lp.\lp. \hphantom{=}
        + \Phi_R^2 + \Phi_I^2\rp)\xi_{B} 
        -4 \lp( \phi_M \p_r \Phi_M + \Phi_M^2 \rp) \D^2\xi_{G}
        \rp.\rp.
        \\ \nonumber
        &\mathopen{}\lp.\lp. \hphantom{=}
        -\frac{8\Lambda_r}{r} + \frac{4\Lambda_t\p_t a}{\alpha} 
        + \frac{4 \Lambda_r \p_r a}{a^3}
        \vphantom{\frac{1}{1}} \rp) \vphantom{\frac{1}{1}} \rp] \zeta,
\end{align}
\begin{align}
        \label{bsgm_l_app_SC_eqn_time}
        \frac{\p_r \alpha}{a^2 r} &= \frac{\alpha \p_r a}{a^3 r}
        +\frac{\alpha}{r^2} - \frac{\alpha}{a^2 r^2} + 
        \lp[\vphantom{\frac{1}{1}} -\alpha V 
        -\frac{\alpha \D^2 \phi_M^2}{r^2}
        \rp.
        \\ \nonumber
        &\hphantom{=}\mathopen{}\lp.
        + \frac{\lp(\phi_R \p_t \Pi_R + \phi_I \p_t \Pi_I\rp)\xi_{B} 
        + \D^2\xi_{G} \phi_M \p_t \Pi_M}{a} 
        \rp.
        \\ \nonumber
        &\hphantom{=}\mathopen{}\lp.\vphantom{\frac{1}{1}}
        +\frac{\alpha \xi_{B}\lp( -\phi_R \p_r \Phi_R + \Pi_R^2 
        - \Phi_R^2 \rp)}{a^2}
        \rp.
        \\ \nonumber
        &\hphantom{=}\mathopen{}\lp.\vphantom{\frac{1}{1}}
        +\frac{\alpha \xi_{B}\lp( -\phi_I \p_r \Phi_I + \Pi_I^2 
        - \Phi_I^2 \rp)}{a^2}
        \rp.
        \\ \nonumber
        &\hphantom{=}\mathopen{}\lp.
        +\frac{\alpha \xi_{G} \D^2 \lp( -\phi_M \p_r \Phi_M + \Pi_M^2 
        - \Phi_M^2 \rp)}{a^2}  
        \rp.
        \\ \nonumber
        &\hphantom{=}\mathopen{}\lp.
        -\frac{4 \Lambda_r \alpha}{a^2 r}
        -\frac{\Lambda_r}{a}\p_r\lp(\frac{\alpha}{a}\rp)
        \vphantom{\frac{1}{1}}\rp]\zeta.
\end{align}

\section{Linear Perturbation Theory Equations
\label{bsgm_l_app_linear_perturbation_theory_equations}}

Following the decomposition of Section~\ref{bsgm_l_sec_lpt}, we write the 
boson star field, $\Psi$, as
\begin{align}
   \Psi &= e^{-i\omega t}\lp(\psi_R+i\psi_I\rp) = \phi_R+i\phi_I,
   \\
   \phi_R(r,t) &= \cos\lp(\omega t\rp)\psi_R(r,t) 
   + \sin\lp(\omega t\rp)\psi_I(r,t),
   \\
   \phi_I(r,t) &= \cos\lp(\omega t\rp)\psi_I(r,t) 
   - \sin\lp(\omega t\rp)\psi_R(r,t),
\end{align}
and perturb about the stationary solutions $\mu_0(r)$, $\nu_0(r)$, $\psi_0(r)$ 
and $\phi_0(r)$:
\begin{align}
   \mu(r,t)&=\mu_0(r) + \epsilon \delta \mu(r) e^{i\beta t}, 
   \\
   \nu(r,t)&=\nu_0(r) + \epsilon \delta \nu(r) e^{i\beta t}, 
   \\
   \psi_R(r,t)&=\psi_0(r) + \epsilon \delta \psi_{R}(r) e^{i\beta t}, 
   \\
   \psi_I(r,t)&=\frac{\epsilon}{i\beta} \delta \psi_{I}(r) e^{i\beta t}, 
   \\
   \phi_M(r,t)&=\phi_0(r) + \epsilon \delta \phi(r) e^{i\beta t}.
\end{align} 
Expanding the equations of motion to first order in $\epsilon$, we find a 
complicated set of equations for linearized non-minimally coupled 
perturbations. To reduce the complexity of these equations, we restrict 
ourselves to the minimally coupled case ($\xi_{\mathrm{B}} = \xi_{G} = 0$) 
whereby the stationary solutions satisfy:
\begin{align}
   \p_r\mu_0 &=
   \lp(\lp(\frac{\lambda_B \psi_0^4}{4} + \lp(\frac{m^2}{2}
   +\frac{\omega^2}{2 e^{\nu_0}}\rp)\psi_0^2 
   \rp.\rp.
   \\ \nonumber
   &\hphantom{=}\mathopen{}\lp.\lp.
   + \frac{\lambda_G\Delta^4\lp(\phi_0^2-1\rp)^2}{4}\rp)e^{\mu_0} 
   + \frac{\Delta^2\lp(\p_r\phi_0\rp)^2}{2} 
   \rp.
   \\ \nonumber
   &\hphantom{=}\mathopen{}\lp.
   + \frac{\lp(\p_r \psi_0\rp)^2}{2} \rp)r 
   + \frac{1+\lp(\Delta^2\phi_0^2-1\rp)e^{\mu_0}}{r},
\end{align}
\begin{align}
   \p_r\nu_0 &=
   \lp(\lp(\frac{-\lambda_B \psi_0^4}{4} - \lp(\frac{m^2}{2}
   -\frac{\omega^2}{2 e^{\nu_0}}\rp)\psi_0^2 
   \rp.\rp.
   \\ \nonumber
   &\hphantom{=}\mathopen{}\lp.\lp.
   - \frac{\lambda_G\Delta^4\lp(\phi_0^2-1\rp)^2}{4}\rp)e^{\mu_0} 
   + \frac{\Delta^2\lp(\p_r\phi_0\rp)^2}{2} 
   \rp.
   \\ \nonumber
   &\hphantom{=}\mathopen{}\lp.
   + \frac{\lp(\p_r \psi_0\rp)^2}{2} \rp)r 
   - \frac{1+\lp(\Delta^2\phi_0^2-1\rp)e^{\mu_0}}{r},
\\
   \p_r^2\psi_0 
   &= \lp(\lambda_B\psi_0^3+\lp(m^2
   -\frac{\omega^2}{e^{\nu_0}}\rp)\psi_0\rp) e^{\mu_0}
   \\ \nonumber
   & \hphantom{=}
   +\lp(\frac{\p_r\mu_0}{2} - \frac{\p_r\nu_0}{2} -\frac{2}{r}\rp)\p_r 
   \psi_0,
\\
   \p_r^2\phi_0 &=
   \lambda_G\Delta^2\lp(\phi_0^3-\phi_0\rp)e^{\mu_0} + \frac{2\phi_0 e^
   {\mu_0}}{r^2}
   \\ \nonumber
   & \hphantom{=}
   + \lp(\frac{\p_r\mu_0}{2} - \frac{\p_r\nu_0}{2} -\frac{2}{r}  \rp)\p_r
   \phi_0,
\end{align}
and the equations for the perturbed quantities reduce to the following:
\begin{align}
   \label{bsgm_l_app_pert_start}
   \p_r {\delta N} &= 
   r^2e^{\frac{1}{2}\lp(\mu_0-\nu_0\rp)}
   \lp(\frac{\lp(\delta \mu - \delta \nu\rp)\omega\psi_0^2}{4}
   + \lp(\vphantom{\frac{1}{1}}\omega \delta \psi_R 
   \rp.\rp.
   \\ \nonumber
   & \hphantom{=} \mathopen{}\lp.\lp.
   - \frac{\delta \psi_I}{2}\rp)\psi_0\rp).
\\
   \p_r \delta \mu &= 
   \lp(\lp(\lambda_G\Delta^4 \lp(\phi_0^3-\phi_0\rp)+\frac{2\Delta^2\phi_0}
   {r^2}\rp)\delta\phi
   \rp.
   \\ \nonumber
   &\hphantom{=}\mathopen{}\lp.
   +\lp(\frac{\lambda_B\psi_0^4 }{4} + \lp(\frac{m^2}{2}+\frac{\omega^2}
   {2e^{\nu_0}}\rp)\psi_0^2
    + \frac{\Delta^2\phi_0^2-1}{r^{2}}
   \rp.\rp.
   \\ \nonumber
   &\hphantom{=}\mathopen{}\lp.\lp.
   +\frac{\lambda_G\Delta^4 \lp(\phi_0^2-1\rp)^2}{4}\rp)\delta \mu
   +\lambda_B\psi_0^3\delta\psi_R - \frac{\omega^2\psi_0^2\delta
   \nu}{2 e^{\nu_0}}
   \rp.
   \\ \nonumber
   &\hphantom{=}\mathopen{}\lp.
   +\lp(\lp(m^2+\frac{\omega^2}{e^{\nu_0}}\rp)\delta\psi_R -\frac{\omega
   \delta\psi_I}{e^{\nu_0}}\rp)\psi_0
   \rp)e^{\mu_0}r
   \\ \nonumber
   & \hphantom{=}
   +r\Delta^2\lp(\p_r \phi_0\rp)\lp(\p_r\delta\phi\rp) + r\lp(\p_r \psi_0
   \rp)\lp(\p_r
   \delta \psi_R\rp),
\\
   \p_r\delta\nu &=
   \lp(\lp(2\lambda_G\Delta^4\lp(\phi_0-\phi_0^3\rp) - \frac{4\Delta^2
   \phi_0}{r^{2}}\rp)
   \delta\phi
   \rp.
   \\ \nonumber
   &\hphantom{=}\mathopen{}\lp.
   -\frac{1}{2}\lp(\lambda_G\Delta^4\lp(\phi_0^2-1\rp)^2 + \frac{4\lp(
   \Delta^2\phi_0^2-1\rp)}{r^2}
   \rp.\rp.
   \\ \nonumber
   &\hphantom{=}\mathopen{}\lp.\lp. \vphantom{\frac{1}{1}}
   + 2m^2\psi_0^2+\lambda_B\psi_0^4\rp)\delta\mu
   -2\lp(\lambda_B\psi_0^3
   \rp.\rp.
   \\ \nonumber
   &\hphantom{=}\mathopen{}\lp.\vphantom{\frac{1}{1}}\lp.
   +m^2\psi_0\rp)\delta\psi_R
   \rp)e^{\mu_0}r + \p_r\delta\mu,
\\
   \p_r^2\delta\psi_R &=
   \frac{1}{2}\lp(\p_r\mu_0-\p_r\nu_0-\frac{4}{r}\rp)\p_r{\delta \psi_{R}}
   \\ \nonumber
   & \hphantom{=}
   +\frac{\lp(\p_r\delta\mu - \p_r \delta\nu \rp)\p_r\psi_0}{2}
   +\lp(\vphantom{\frac{1}{1}} \frac{\lp(\delta\nu-\delta\mu\rp)\omega^2\psi_0}
   {e^{\nu_0}}
   \rp.
   \\ \nonumber
   & \hphantom{=}\mathopen{}\lp.
   +\lp(m^2+3\lambda_B\psi_0^2\rp)\delta\psi_R
   + \lp(m^2\psi_0+\lambda_B\psi_0^3\rp)\delta\mu
   \rp.
   \\ \nonumber
   & \hphantom{=}\mathopen{}\lp.
   + \frac{2\omega\delta\psi_I-\lp(\omega^2+\beta^2\rp)\delta\psi_R}
   {e^{\nu_0}} \rp)e^{\mu_0},
\end{align}
\begin{align}
   \p_r^2\delta\psi_I &=
   \frac{1}{2}\lp(\p_r\mu_0-\p_r\nu_0-\frac{4}{r}\rp)\p_r{\delta \psi_{I}}
   \\ \nonumber
   & \hphantom{=}
   +\lp(\lp(m^2+\lambda_B\psi_0^2 - \frac{\lp(\omega^2+\beta^2\rp)}{e^
   {\nu_0}}\rp)\delta\psi_I
   \rp.
   \\ \nonumber
   & \hphantom{=}\mathopen{}\lp.
   + \frac{\omega\beta^2\lp(\delta\mu-\delta\nu\rp)\psi_0}{2 e^{\nu_0}} 
   + \frac{2\omega\beta^2 \delta\psi_R}{e^{\nu_0}}\rp)e^{\mu_0},
\\
   \label{bsgm_l_app_pert_end}
   \p_r^2\delta\phi &=
   \frac{1}{2}\lp(\p_r\mu_0-\p_r\nu_0-\frac{4}{r}\rp)\p_r{\delta \phi}
   \\ \nonumber
   & \hphantom{=}
   + \lp( \lp( \lambda_G\Delta^2\lp(3\phi_0^2-1\rp) + \frac{2}{r^2} \rp)
   \delta\phi
   \rp.
   \\ \nonumber
   &\hphantom{=}\mathopen{}\lp.
   +\lp(\lambda_G\Delta^2\lp(\phi_0^3-\phi_0\rp) + \frac{2\phi_0}{r^2}\rp)
   \delta\mu
   \rp.
   \\ \nonumber
   &\hphantom{=}\mathopen{}\lp.
   -\frac{\beta^2\delta\phi}{e^{\nu_0}}
   \rp)e^{\mu_0} +\frac{\lp(\p_r\delta\mu - \p_r \delta\nu\rp)\p_r\phi_0}{2},
\end{align}

Given that the deviations from the stationary solutions given by 
(\ref{bsgm_l_app_pert_start}--\ref{bsgm_l_app_pert_end}) involve 
perturbations of three dynamic fields (with the metric perturbations having 
no dynamic freedom of their own), it is not immediately obvious that the 
perturbations should be purely exponential or oscillatory in time. 
For example, one could imagine perturbations involving under or over damped 
oscillations corresponding to complex $\beta^2$. Here, we follow the work of 
Jetzer~\cite{jetzer1992boson}, and demonstrate that a set of equations 
equivalent to (\ref{bsgm_l_app_pert_start}--\ref{bsgm_l_app_pert_end}) 
may be written in the following form,
\begin{align}
   \mathrm{L}_{ij}\mathbf{f}_{j}=-\beta^2e^{\mu_0-\nu_0}\mathrm{G}_{ij}
   \mathbf{f}_j,
\end{align}
where $\mathrm{L}_{ij}$ is a Hermitian differential operator, 
$\mathbf{f_j}$ is the solution vector and $\mathrm{G}_{ij}$ is a diagonal 
matrix. It then follows that the eigenvalues of the above pulsation equation 
(for $\beta^2$) are purely real. As before we expand our fields about the 
stationary solutions, but do not yet enforce exponential time dependence:
\begin{align}
   \mu(t,r)&=\mu_0(r) + \epsilon \delta \mu(t,r), 
   \\
   \nu(t,r)&=\nu_0(r) + \epsilon \delta \nu(t,r), 
   \\
   \psi_R(t,r)&=\psi_0(r) + \epsilon\delta\psi_R(t,r), 
   \\
   \psi_I(t,r)&=\epsilon \psi_0(r) \delta\psi_I(t,r), 
   \\
   \phi(t,r)&=\phi_0(r) + \epsilon \delta \phi(t,r).
\end{align} 

Substituting these expressions into the equations of motion for the fields 
and metric and truncating to linear order in the perturbation, we find a 
coupled system of perturbation equations. In particular, the equation 
$G_{rt}=T_{rt}$ produces the following simple expression for $\delta \mu$:
\begin{align}
   \label{bsgm_l_app_dmu_herm_temp}
   \p_t \delta\mu &= r \lp( \Delta^2 \lp(\p_t {\delta \phi} \rp) 
   \lp( \p_r \phi_0 \rp) - \lp( \p_r {\delta\psi_I} \rp)\psi_0^2\omega
   \rp.
   \\ \nonumber
   &\hphantom{=}\mathopen{}\lp.
   +\lp(\p_r\psi_0\rp)\lp(\p_t\psi_R\rp) \rp).
\end{align}
With the substitution
\begin{align}
   \delta\psi_I=\p_t{\delta \tilde{\psi}_I},
\end{align}
(\ref{bsgm_l_app_dmu_herm_temp}) becomes a total derivative with respect to 
time and may be integrated to give,
\begin{align}
   \label{bsgm_l_app_dmu_herm}
   \delta\mu &= r\lp(\Delta^2\lp(\delta\phi\rp)\lp(\p_r\phi_0\rp) 
   + \lp(\delta\psi_R\rp)\lp(\p_r\psi_0\rp)
   \vphantom{\frac{}{}}\rp.
   \\ \nonumber
   &\hphantom{=}\mathopen{}\lp.
   - \lp(\omega\psi_0^2\rp)\p_r {\delta\tilde{\psi}_I}\rp).
\end{align}
We may also obtain an expression for $\delta \nu$ by solving the perturbed 
$G_{tt}=T_{tt}$ equation for $\delta \nu$ and substituting the relevant 
expressions for $\delta \psi_I$, $\delta \mu$, $\p_r^2 \psi_0$, $\p_r^2 
\phi_0$, $\p_r \mu_0$, and $\p_r \nu_0$:
\begin{align}
   \delta\nu &=
   -\frac{r e^{\nu_0}\lambda_G\Delta^4\lp(\phi_0^2-1\rp)^2
   \p_r {\delta \tilde{\psi}_I}}{2\omega} -\frac{2\p_t^2
   \delta\tilde{\psi}_I}{\omega}
   \\ \nonumber
   & \hphantom{=}
   +\Delta^2\lp( \delta\phi\lp( \p_r \phi_0 \rp)r 
   - \frac{2\phi_0^2 e^{\nu_0} \lp(\p_r \delta 
   \tilde{\psi}_I\rp)}{r \omega}  \rp)
   \\ \nonumber
   & \hphantom{=}
   +e^{\nu_0}\lp(\frac{2}{r\omega} - \frac{\lp(\lambda_B\psi_0^4 
   +2m^2\psi_0^2\rp)r}{2\omega} \rp) \p_r { \delta \tilde{\psi}_I }
   \\ \nonumber
   & \hphantom{=}
   -r \omega \psi_0^2 \lp(\p_r\delta\tilde{\psi}_I\rp) 
   +\lp(r\p_r\psi_0 +\frac{4}{\psi_0}  \rp)\delta \psi_R
   \\ \nonumber
   & \hphantom{=} 
   +2\lp(\lp(\frac{2\p_r\psi_0}{\psi_0} + \frac{1}{r}\rp)
   \p_r{\delta\tilde{\psi}_I} + \p_r^2\delta\tilde{\psi}_I\rp)
   \frac{e^{\nu_0}}{\omega e^{\mu_0}}.
\end{align}

With the perturbed metric functions now defined solely in terms of the 
stationary solution and perturbed matter fields, we find expressions for 
$\p_r^2 \delta \psi_R$, $\p_r^3 \delta\tilde{\psi}_I$  and $\p_r^2 \delta 
\phi$ through substitution. At this stage, the perturbation equations 
consist of three second order expression in $\mathbf{f}_j=( \delta \psi_R, 
\p_r \delta \tilde{\psi}_I, \delta \phi)$ with all time derivatives 
appearing as second order expressions. Substituting:
\begin{align}
   \delta \tilde{\psi}_I(r,t) = \delta \tilde{\psi}_I(r) e^{i\beta t}, 
\\
   \delta \psi_R(r,t) = \delta \psi_R(r) e^{i\beta t},
\\
   \delta \phi(r,t) = \delta \phi(r) e^{i\beta t},
\end{align}
we may write our coupled perturbation expressions as,
\begin{align}
   \mathrm{\tilde{L}}_{ij}\mathbf{f}_{j}=-\beta^2e^{\mu_0-\nu_0} \mathbf{f}_j,
\end{align}
where $\mathrm{\tilde{L}}_{ij}$ is a second order, non-Hermitian operator 
of the following form:
\begin{align}
   \left[
   \begin{matrix}
      \frac{\partial^2}{\partial r^2} + a_1\frac{\partial}{\partial r} 
      + b_1 & c_1\frac{\partial}{\partial r} +d_1 & e_1 \\
      c_2\frac{\partial}{\partial r} +d_2 &  \frac{\partial^2}{\partial r^2}
      + a_2\frac{\partial}{\partial r} + b_2 & h_2 \\
      e_3 & h_3 & \frac{\partial^2}{\partial r^2} + a_3\frac{\partial}
      {\partial r} + b_3 \\
   \end{matrix}
   \right].
\end{align}

Here, all subscripted quantities should be understood to 
be functions of $r$.
The goal is now to find a diagonal matrix, $\mathrm{M}_{ij}$, 
such that $\mathrm{L}_{ij} =  \mathrm{M}_{ik} \mathrm{\tilde{L}}_{kj}$ is a 
Hermitian operator of the form\begin{align}
   \left[
   \begin{matrix}
      \frac{\partial}{\partial r}M_1\frac{\partial}{\partial r} 
      + B_1 & -\frac{\partial}{\partial r}C_1 
      + D_1 & E_1 
      \\
      C_1\frac{\partial}{\partial r} +D_1 & \frac{\partial}
      {\partial r}M_2\frac{\partial}{\partial r} + B_2 
      & H_1
      \\
      E_1 & H_1 & \frac{\partial}{\partial r}M_3
      \frac{\partial}{\partial r} + B_3 \\
   \end{matrix}
   \right],
\end{align}
and
\begin{align}
   \mathrm{M}_{ij}=
   \left[
   \begin{matrix}
      M_1 & 0 & 0 \\
      0 & M_2 & 0 \\
      0 & 0 & M_3 \\
   \end{matrix}
\right].
\end{align} 
Fortunately, this turns out to be a well defined problem and the $M_i$'s and 
$C_1$ take the form:
\begin{align}
   M_1 &= r^2e^{\frac{1}{2}\left(\nu_0-\mu_0\right)}, \\
   M_2 &= r^2e^{\frac{3}{2}\left(\nu_0-\mu_0\right)}\psi_0^2, \\
   M_3 &= r^2\Delta^2 e^{\frac{1}{2}\left(\nu_0-\mu_0\right)}, \\
   C_1 &= 2 \omega r^2 \psi_0 e^{\frac{1}{2}\lp(\nu_0-\mu_0\rp)},
\end{align}
while the remaining terms are sufficiently cumbersome that it is not 
particularly enlightening to write them out explicitly. One could, of 
course, use the equations just derived to solve the perturbation problem 
rather than~(\ref{bsgm_l_app_pert_start}--\ref{bsgm_l_app_pert_end}). 
Unfortunately, as is often the case with such matters, by the time we 
verified that the equations permitted only real eigenvalues, the previous 
formalism had already been adopted and investigated. Due to the significant 
reduction in complexity these equations represent, we would highly recommend 
future work to follow this approach rather than the more direct method 
we adopted.

\bibliography{bsgm_stability}

\end{document}